\documentclass[11pt]{article}

\usepackage{graphicx}
\usepackage{amsmath}
\usepackage{xcolor}
\usepackage{ragged2e}
\usepackage{amssymb}
\usepackage{tabularx}
\usepackage{booktabs}
\usepackage{subfig}
\usepackage{empheq}
\usepackage{stackrel}
\usepackage[font={small,it}]{caption}
\usepackage{hyperref}
\usepackage{url}
\usepackage{textcomp}
\usepackage{upgreek}
\usepackage{setspace}
\usepackage{comment}
\usepackage{cleveref}
\usepackage{float}
\usepackage{multirow}

\definecolor{my_green}{rgb}{0.09, 0.45, 0.27}
\definecolor{my_blue}{rgb}{0.01, 0.28, 1.0}
\definecolor{mygrey}{rgb}{0.6, 0.6, 0.6}

\newcommand{\red}[1]{\textcolor{red}{#1}}

\title{\bf Analyzing cell-to-cell heterogeneities and cell configurations in parallel-connected battery modules using physics-based modeling}
\author{Simone Fasolato\thanks{ Department of Electrical, Computer and Biomedical Engineering, University of Pavia, Pavia, Italy.} , Anirudh Allam$^{\ddagger}$,  Simona Onori\thanks{Energy Science and Engineering, Stanford University, Stanford, CA. \\ Corresponding author  {\tt\small simone.fasolato01@universitadipavia.it}}, and Davide M. Raimondo\thanks{ Department of Engineering and Architecture, University of Trieste, Italy.} } 
\date{}

\begin{document}
\maketitle	

\setstretch{1}
\setcounter{MaxMatrixCols}{13}

\vspace{-1em}
\begin{abstract}
\noindent
In parallel-connected cells, cell-to-cell (CtC) heterogeneities 
can lead to current and thermal gradients that 
may adversely impact the battery performance and aging. Sources of CtC heterogeneity include manufacturing process tolerances, poor module configurations, and inadequate thermal management. 
Understanding which CtC heterogeneity sources most significantly impact battery performance is crucial, as it can provide valuable insights.
In this study, we use an experimentally validated electrochemical battery model to simulate hundreds of battery configurations, each consisting of four cells in parallel. We conduct a statistical analysis to evaluate the relative importance of key cell-level parameters, interconnection resistance, cell spacing, and location on performance and aging.
 The analysis reveals that heterogeneities in electrode active material volume fractions primarily impact module capacity, energy, and cell current, leading to substantial thermal gradients. However, to fully capture the output behavior, interconnection resistance, state of charge gradients and the effect of the temperature on parameter values must also be considered.
  Additionally, module design configurations, particularly cell location, exacerbate 
 thermal gradients, accelerating long-term module degradation. This study also offers insights into optimizing cell arrangement 
 during module design to reduce thermal gradients and enhance overall battery performance and longevity. 
 Simulation results 
 with four cells 
 indicate a reduction of 51.8\% in thermal gradients, leading to a 5.2\% decrease in long-term energy loss.
\end{abstract}

\section{Introduction} 
A critical challenge towards accelerating the energy transition from fossil fuels to renewable energy resources is the development of reliable, long-lasting, and safe energy storage systems. 
Lithium-ion batteries are widely recognized  as the dominant energy storage technology for applications in portable electronics,  automotive industry \cite{muratori2021rise}, and renewable energy \cite{choi2021li}.
To meet precise power and energy demands while ensuring optimal performance and safe operation, lithium-ion battery modules or packs, consisting of interconnected individual cells in series and/or parallel arrangements, are combined with battery management systems (BMS) \cite{rahimi2013battery}.
The BMS enables the estimation of unmeasurable cell states such as State of Charge (SOC) and State of Health (SOH), as well as, it is responisble for cell balancing and fault detection strategies, ensuring each cell operates optimally. 
Additionally, the BMS is tasked with the thermal management control, ensuring the battery pack operates within safe temperature limits.
However, a crucial aspect in enhancing the BMS control and estimation algorithms is acknowledging the presence of cell-to-cell (CtC) heterogeneity and understanding their influence on pack/module performance, degradation, and safety \cite{schuster2015lithium}.

\paragraph{CtC variations causes}
$\\$
Prior research has effectively identified and synthesized the key factors that significantly impact the efficacy of battery modules \cite{baumann2018parameter}. 
Heterogeneities in batch of fresh cells are typically attributed to manufacturing tolerances during production processes and/or differences in material composition \cite{rumpf2017experimental}. Manufacturing-related CtC variations can manifest as variations in internal resistance \cite{gogoana2014internal,reiter2019holistic,shi2016effects}, capacity \cite{miyatake2013discharge,zhang2015study}, or a combination of both \cite{bruen2016modelling, pastor2016study,cordoba2015control}.
Examples of distributions of cell characteristics outside of manufacturer specifications can be found in \cite{rumpf2017experimental} and \cite{campestrini2016ageing, xie2020facile} for LFP/graphite and NCA/graphite fresh cell batches, respectively.
Not only single-cell level features but also module-level characteristics strongly contribute to introduced CtC variation. 
According to \cite{baumann2018parameter}, electrical resistance among the cell interconnections \cite{jocher2021novel,li2022effect} stands as the second leading factor contributing to heterogeneity in a battery module. 
Typically, variations in interconnection resistances are caused by factors such as weld cracks \cite{brand2015welding}, faulty connections between cells and busbars \cite{ brand2017electrical}, contact imperfections among electrodes and current collectors due to material irregularities and uneven pressure \cite{taheri2011investigating}, and improperly sized electrical connections that lead to increased local resistance \cite{fill2020algorithm}.
Additionally, the number of cells in parallel \cite{fill2019analytical,diao2019management}, topology selection \cite{luan2021influence,grun2018influence}, and chemistry combination \cite{chang2022experimental,tian2022parallel} have a non-negligible impact on pack performance. 
Finally, operating temperature \cite{yang2016unbalanced, wu2013coupled} and poor cooling design-induced thermal gradients \cite{naylor2023battery,klein2017current} can also affect the uniformity of pack performance.

\paragraph{CtC variations effects}
$\\$ The effect of CtC heterogeneity on the operation of a battery module or pack varies based on the interconnection configuration, whether connected in series or parallel.

In the \textbf{series-connected cells} scenario, the overall capacity of the module is constrained by the weakest cell \cite{kenney2012modelling}, which is the one with the lowest capacity. Additionally, the module degradation is accelerated by thermal gradients among the cells \cite{allam2019exploring}, resulting from uneven heat generation due to heterogeneous cell internal and/or interconnection resistances, as well as a suboptimal cooling system.
On the other hand, in a \textbf{parallel-connected module}, dissimilar cell capacities, resistances, and temperatures result in heterogeneous current distribution \cite{weng2022parallel}, which in turn leads to thermal and SOC imbalances \cite{brand2016current}. This results in cell-to-cell fluctuations in internal resistance, capacity \cite{campestrini2016ageing, schindler2021analyzing, naylor2023battery}, and aging rate \cite{wang2019dependency,baumann2018parameter, an2016cell} over time.
In particular, the phenomenon of performance imbalance leading to non-uniform aging of individual cells has been reported in the literature. Specifically, it has been noted that the prolongation of imbalances is a contributing factor to this phenomenon. The existing literature presents divergent views on this matter. While some researchers \cite{pastor2016study, bruen2016modelling, spurrett2002modeling, brand2012ageing} affirm that there exists a convergence and self-balancing tendency among parallel-connected cells over time, others' \cite{schuster2015lithium, gong2014study, cordoba2015control, kakimoto2015capacity} findings oppose this theory. So far, most research has focused on individual cells' behavior, with some experimental assessments of module connections reported in \cite{baumann2018parameter, jocher2021novel}. Consequently, the issue of parallel cell connections leaves gaps in the knowledge and necessitates further investigation.

\vspace{+1em}
In parallel-connected battery modules CtC variations cells are closely interdependent and, in practical terms, challenging to prevent. Due to the absence of individual cell sensors, these modules are managed as single lumped cells within a BMS, ignoring internal heterogeneity. 
This oversight can lead to undetected current imbalances, resulting in cell overcharging or overdischarging, thermal gradients, hotspots, potential safety hazards, and distinct degradation rates among cells. 
While systematic testing at module and pack levels could provide valuable insights, it is economically and practically unfeasible due to high costs and material waste. 
Considering the multitude of factors involved in module and pack design, along with the long-term effects of CtC variations on battery system aging, utilizing digital twins presents an effective alternative approach \cite{li2020digital}. 
Digital twins employ high-fidelity mathematical models that capture CtC interactions and replicate both cell heterogeneity and module responses. This methodology provides valuable insights into how heterogeneity propagates within parallel battery modules, especially under stringent economic, temporal, and facility constraints that hinder the execution of complex experimental campaigns. 

The main objective of this study is to investigate the impact of CtC variation on parallel-connected battery modules through a model-based statistical approach.
The study introduces a comprehensive experimentally-validated physics-based modeling framework. 
In this framework, the electrochemical dynamics of each cell are modeled using the Enhanced Single Particle Model (ESPM)\cite{tanim2015temperature},  coupled with a thermal model that considers the thermal interconnection among cells and physics-based cell aging model.
SCompared to several Equivalent Circuit Model (ECM) based parallel-connected modules proposed in the literature \cite{gong2014study, brand2016current,bruen2016modelling, hosseinzadeh2021quantifying, klein2017current, piombo2021analysis, gogoana2014internal, fan2020simplified, cordoba2015control} , the present model accurately monitors the electrochemical states of each cell and is better suited for offline high-fidelity simulations.
Additionally, the ESPM is preferred over more complex DFN-based module models \cite{wu2013coupled, tian2022parallel, rumpf2018influence, li2022effect}  because it offers comparable accuracy at mild C-rates \cite{lee2021robust} while substantially reducing computational requirements for large-scale and long-duration simulations.

Leveraging the experimentally validated electrochemical-thermal-aging model at the module level, the main contributions of this paper are:
\begin{enumerate}
  \item A statistical analysis, based on a multi-linear regression (MLR) approach \cite{tonidandel2011relative}, that evaluates how CtC variations, such as capacity-related heterogeneities, cell spacing, interconnection resistance, and cell location, affect overall module capacity, energy, current, and thermal distribution, as well as aging propagation. By performing this analysis, we aim to gain a deeper understanding of how these factors impact the performance and reliability of parallel-connected battery modules. Our analysis reveals that heterogeneities in electrode active material volume fractions primarily impact module capacity, energy, and cell current, leading to substantial thermal gradients. However, to fully capture the output behavior, interconnection resistance, state of charge gradients, and the effect of temperature on parameter values must also be considered.
  \item A simple cell arrangement strategy for parallel-connected battery modules. The key idea is to leverage the heterogeneous cell current distribution to mitigate the module's thermal heterogeneity and overall long-term degradation. 
  Note that, in the literature, optimized cell arrangement has been highlighted as crucial for reducing thermal gradients within the  module casing and for improving uniformity in cell aging. Among notable  works, the authors in \cite{chen2018design} introduced a non-uniform cell spacing strategy that lowered maximum temperature by $3^\circ\mathrm{C}$ and decreased thermal gradient by 60\% compared to a conventional uniform layout. Furthermore, \cite{saechan2022numerical} compares inline, offset, and staggered module configurations, noting that the offset arrangement decreases the power consumption of the battery thermal  management system by an impressive 43.1\% relative to the other configurations.
\end{enumerate}

This paper is structured as follows: Section~\ref{Sec:Method} presents the materials and methods used in this research, particularly the experimental campaign, the detailed identification and validation of the high-fidelity model, and the basis for constructing the MLR model. Section~\ref{Sec:Res_and_disc} discusses the outcomes of the MLR-based analysis regurading the impacts of CtC variations on the parallel-connected module, and introduces the proposed cell arrangement strategy, and Section~\ref{Sec:Conclusion} summarizes the main conclusions of this work.

\section{Material and methods} \label{Sec:Method}
The structure of this section is outlined as follows: Section \ref{Sec:Battery_testing} details both cell-level and module-level experiments necessary for model identification and validation. In Sections \ref{Sec:Met_Model} and \ref{Sec:Model_validation}, we introduce the mathematical model developed for the parallel-connected module and describe the procedures for model validation, respectively. Finally, Section \ref{Sec:MLR} details the multi-linear regression approach used for statistical analysis.

\subsection{Battery testing} \label{Sec:Battery_testing}
In this section, we provide a comprehensive overview of the battery testing procedures implemented at both the cell and module levels. 

It is woth noting that, to provide the desired current profiles to single cells and modules and collect sensor data (i.e. voltage, Hall sensor voltages, and cell surface temperatures) the Arbin LBT21024 and Arbin LBT22013 cyclers are used, respectively.
Each test is conducted in an Amerex IC500R thermal chamber, and every cell is equipped with a T-type thermocouple at its center to measure surface temperatures. Additionally, during module-level testing, four Honeywell SS495A Hall sensors are installed in each module to monitor currents in parallel branches. These Hall sensors are inserted and glued into ferrite rings to enhance the signal-to-noise ratio, further, shielded cables are employed to enclose the signals during operation.
For more detailed information on the single-cell and module testing campaigns, please refer to references \cite{piombo2024full} and \cite{piombo2024unveiling}.

\begin{figure}
  \centering
  \includegraphics[width = 0.5\textwidth]{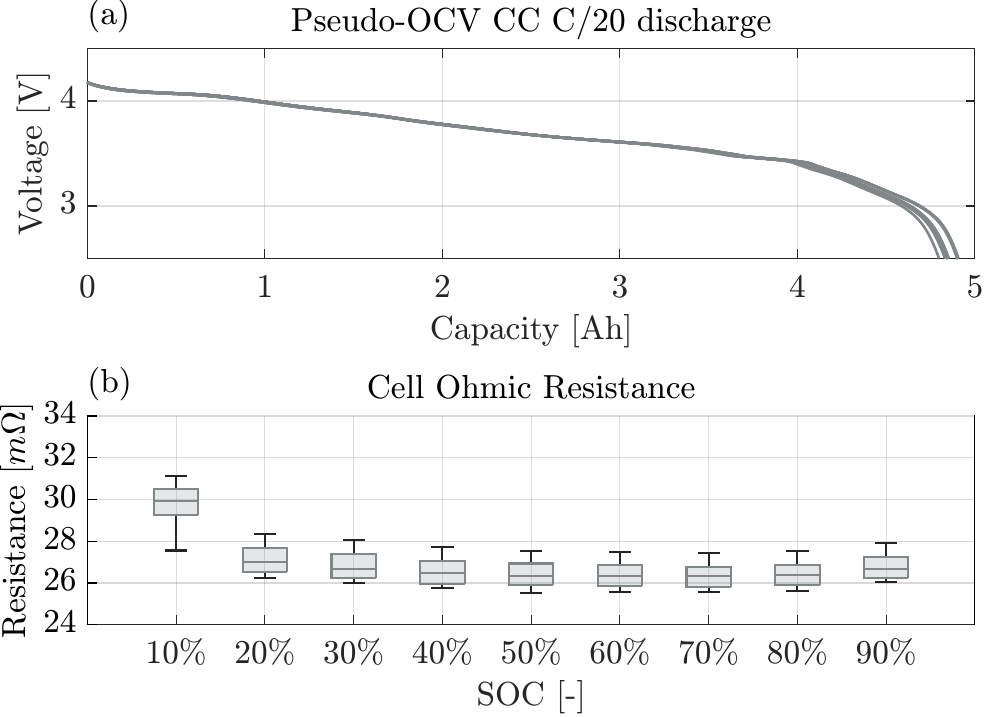}
  \caption{ (a) Pseudo OCV curves of the fresh 19 cells under C/20 discharge test procedure at 23 $^{\circ} C$. (b) Boxplot of cells ohmic resistances at 10\% SoC intervals at 23 $^{\circ} C$}.
  \label{fig:LG_cap_res}
\end{figure}

\subsubsection{Single cell characterization} \label{Sec:Cell_testing}
The LG Chem INR21700 M50T cells are constructed with a negative electrode made of silicon-doped graphite (SiC) and a positive electrode comprising Nickel Manganese Cobalt (NMC) 811 oxide \cite{LGChemM50T}. To characterize the 19 fresh cells, the testing protocol was divided into two steps: a Pseudo-OCV test and a Hybrid Pulse Power Characterization (HPPC) test augmented by a Multi-Sine (MS) procedure \cite{widanage2016design}. Both steps were performed at 23 $^{\circ} C$. 
Initial conditioning of all tests included a constant current-constant voltage (CCCV) charging phase at a C/3 rate until the charging current dropped to 50 mA at a voltage threshold of 4.2 V. The Pseudo-OCV involved discharging the cells at a constant current of C/20 until the voltage reaches the cut-off limit of 2.5 V.
For the combined HPPC and MS test, the aim was to investigate the impact of the SOC on the internal properties of the cells. This was achieved through dynamic current profiles applied at regular SOC intervals of 10\%, each preceded by a 1C rate discharge and a 60-minute rest period. At each predetermined SOC, an HPPC pulse with a charge/discharge ratio of 0.75 and a pulse duration of 10 seconds was utilized \cite{christophersen2015battery}. This was followed by MS dynamic current profiles adhering to the methodology outlined in \cite{widanage2016design}, utilizing an alpha ($\alpha$) value of 0.6 and a pulse duration of 10 seconds.

The experimental campaign on individual cells is designed to characterize the distribution of internal features once they have completed the manufacturing process, as illustrated in Figure \ref{fig:LG_cap_res}. Specifically, a C/20 CC discharge test is employed to evaluate heterogeneities in cell capacities ($Q_{cell}$), which are calculated by integrating the discharging current throughout the entire cycle \footnote{$Q_{cell} = \frac{1}{3600} \int_{0}^{t_{end}^{dis}} I_{cell} dt$, as outlined in \cite{ha2023electrochemical}}. For the batch of 19 LG M50T cells, the mean value and standard deviation of $Q_{cell}$ are 4.86 Ah and 0.033 Ah, respectively.
On the other hand, the distribution of cell ohmic resistance at various SOC is assessed through the HPPC cycle, as shown in Figure \ref{fig:LG_cap_res}(b). It is important to note that the cell ohmic resistance is calculated using the ratio of the voltage drop to the current drop immediately following the current pulse, as described in \cite{ha2023electrochemical}.
The collected data are utilized to identify and validate the physics-based ESPMs as reported in Section \ref{Sec:Cell_identification}, obtaining further the model parameter distributions across the 19 cells.

\begin{figure}
  \centering
  \includegraphics[width = 0.97\textwidth]{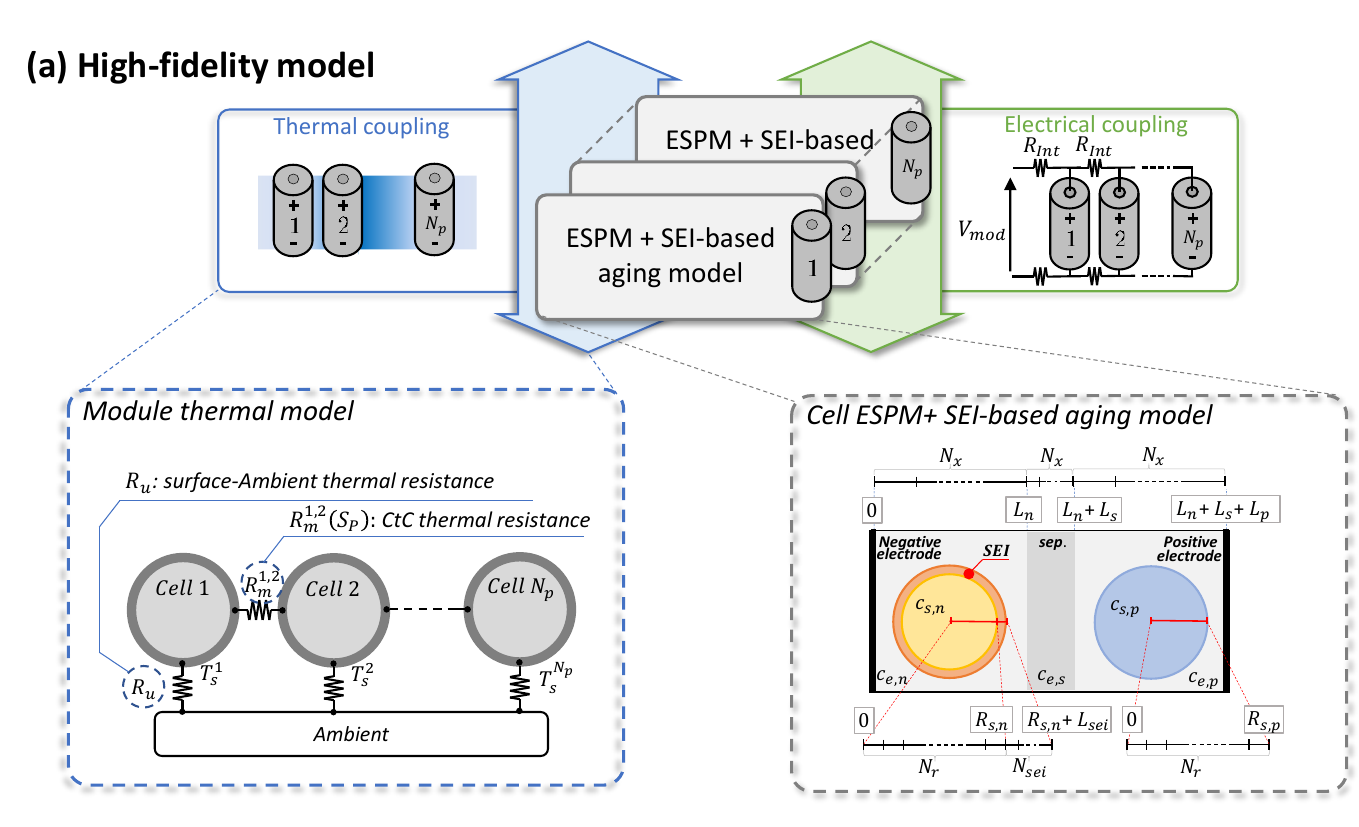}
  \caption{Schematic representation of the physics-based electrochemical-aging-thermal model for the battery module, where the module thermal model and the cell-level electrichemical model are highlighted. }
  \label{fig:module_model}
\end{figure}

\subsubsection{Module-level testing} \label{Sec:Module_testing}
In this work,  battery modules comprising four parallel-connected cells in a ladder configuration, meaning that the module terminals are positioned on the same side as schematically shown in Figure \ref{fig:module_model}, are tested. 
As described in \cite{piombo2024full}, the experiments were carried out under different configurations of interconnection resistance ($R_{int}$) and ambient temperature using an Arbin LBT22013 as a cycler. 
It is worth highlighting that Figure \ref{fig:module_model} schematically reports the placement of $R_{int}$ within each module.
In each tested configuration, the module consisted of four randomly selected NMC LG INR21700 M50T cells from the batch of 19 characterized cells (Section \ref{Sec:Cell_testing}). Then, the module undergoes an initial charging process to achieve 100\% of SOC through a CC-CV cycle, followed by a CC discharge at a rate of 0.75C until reaching 2.5 V, which corresponds to 0\% SOC.
Throughout each test, the overall module current and voltage were monitored, along with the currents supplied by each cell within the module and their respective temperatures, measured with hall sensors and T-type thermocouples, respectively. 
This comprehensive monitoring aimed to track both the current and thermal distribution over the entire discharging cycle.
In Section \ref{Sec:module_model_validation}, the measured quantities are then compared with those obtained from the module-level model for validation purposes.

\subsection{Mathematical model of the parallel-connected module} \label{Sec:Met_Model}
In this section, we present a comprehensive model for a battery module consisting of $N_p$ parallel-connected cells. A schematic representation of the overall framework is illustrated in Figure \ref{fig:module_model}.

The electrochemical dynamics of each cell within the module are modeled using Enhanced Electrolyte Single Particle Models (ESPMs) \cite{tanim2015temperature}. Each ESPM is further integrated with a lumped thermal model \cite{lin2014lumped} and an aging model to accurately capture heat generation within the cell casing and the growth of the Solid Electrolyte Interphase (SEI) layer on the cell negative electrode \cite{safari2008multimodal, prada2013simplified}. Detailed descriptions of the cell-level electrochemical, thermal, and aging models are provided in \textbf{Supplementary Materials Notes 1, 2, and 3}, respectively.

To accurately represent the module behavior, the module-level model builds upon the cell-level model by integrating both electrical and thermal interactions among the cells, ensuring a comprehensive simulation of the module overall performance. The details of both module-level electrocal and thermal models are discussed in the following sections.





\subsubsection{Module-level electrical model} \label{SubSec:ModuleElectrical}
The electrical dynamics of cells connected in parallel, as schematically depicted in Figure \ref{fig:module_model}, are governed by Kirchhoff's circuit laws. Specifically, the current delivered by each cell is determined by solving a system of $N_p$ algebraic equations, represented as:
\begin{equation}
	\label{eq:parallel_pack} 
	\begin{cases}
	V_{cell}^{[k+1]} = V_{cell}^{[k]} + 2R_{int}(I_{mod}-\sum_{z=1}^{k} I_{cell}^{[z]}) \\
	I_{mod} = \sum_{k=1}^{N_{p}} I_{cell}^{[k]} \end{cases}
\end{equation}
where $R_{int}$ represents the interconnection resistance, $I_{mod}$ is the overall input current of the module, and $I_{cell}^{[k]}$ denotes the current delivered by the $k$-th cell. The voltage generated by the $k$-th cell \footnote{The detailed formulation is provided in \textbf{Supplementary Materials Notes 1}}, $V_{cell}^{[k]}$, is defined as:
\begin{equation}
  V_{cell}^{[k]} = U_p^{[k]} +\eta_p^{[k]} -U_n^{[k]} -\eta_n^{[k]} +\Delta\Phi_e^{[k]} - I_{cell}^{[k]}(R_{cell}^{[k]} +R_{\text{SEI}}^{[k]})  
\end{equation}
Here, $U_j$ and $\eta_j$ denote the electrode open-circuit potential and overpotential, respectively, while $\Delta\Phi_e^{[k]}$ represents the electrolyte overpotential. These potentials are calculated based on the lithium-ion concentration within the electrode and electrolyte, as extensively detailed in \textbf{Supplementary Note 1}.
It is important to note that any heterogeneities in terms of $U_j$, $\eta_j$, or $\Delta\Phi_e$ within the module, as well as differences in cell resistances, result in uneven currents flowing through the different branches to satisfy eq. \eqref{eq:parallel_pack} at any given time.


\subsubsection{Module-level thermal submodel} \label{SubSec:ModuleThermal} 
In Figure \ref{fig:module_model}, a schematic representation of the one-dimensional module is depicted, emphasizing the thermal interconnection terms. The single cell model, described in \textbf{Supplementary Note 2}, is modified to incorporate the CtC thermal interconnection. The thermal dynamics of the k-th cell in the module ($T_{cell}^{[k]}$ with $k=1,2, \cdots, N_p$) are formulated to account for the influence of the upstream and downstream cells, as given by:
\begin{equation}
\label{eq:Module_Tsurf}
C_{s} \frac{dT_{cell}^{[k]}}{dt} = I_{cell}^{[k]}(V_{OCV}^{[k]}-V_{cell}^{[k]}) +T_{cell}^{[k]} I_{cell}^{[k]} \frac{dV_{OCP}}{dT_{cell}} + \frac{T_{amb}-T_{cell}^{[k]}}{R_u} - \frac{T_{cell}^{[k]} -T_{cell}^{[k+1]}}{R_m} - \frac{T_{cell}^{[k]} -T_{cell}^{[k-1]}}{R_m}
\end{equation}
where $T_{cell}^{[k-1]}$ and $T_{cell}^{[k+1]}$ denote the surface temperature of the preceding cell ($k-1$) and the following cell ($k+1$) in the module, respectively. The equation considers the convective thermal resistance between the surface and the ambient ($R_u$), and the thermal resistances between adjacent cell surfaces ($R_m$).

It is worth noting that the CtC heat transfer, characterized by the thermal resistance \( R_m \), is dependent on the module design.
According to \cite{lopez2015experimental}, the heat exchange between adjacent cells, in a module composed of cylindrical cells, primarily occurs through two dominant mechanisms: (1) conduction via the interconnection tabs (\( R_m^{\text{tabs}} \)) and (2) conduction through the air (\( R_m^{\text{air}} \)), while neglecting radiative heat transfer.
These thermal resistances are defined as follows \cite{lopez2015experimental}:
\begin{equation} \label{eq:Rm_air}
  R_m^{\text{air}} = \frac{1}{S_{\text{cell}} \, k_{\text{air}}}
\end{equation}
\begin{equation} \label{eq:Rm_tabs}
  R_m^{\text{tabs}} = \frac{w}{A_{\text{cell}} \, k_{\text{tabs}}}
\end{equation}
where $k_{\text{tabs}}$ is the thermal conductivity of the copper tabs, $A_{\text{cell}}$ is the cross-sectional area of the cell, and $w = d + S_p$ is the tab length, calculated as the sum of the cell diameter ($d$) and cell spacing ($S_p$). Additionally, $k_{\text{air}}$ represents the thermal conductivity of air, and $S_{\text{cell}}$ is the shape factor for cylindrical cells, given by:
\begin{equation*}
  S_{\text{cell}} = \frac{2 \pi h}{\cosh^{-1} \left( \frac{4w^2 - 2d^2}{2d^2} \right)}
\end{equation*}
The overall thermal resistance ($R_m$) is determined by combining $R_m^{\text{tabs}}$ and $R_m^{\text{air}}$ in parallel:
\begin{equation}
    R_m = \left( \frac{1}{R_m^{\text{air}}} + \frac{1}{R_m^{\text{tabs}}} \right)^{-1}
\end{equation}
It is important to note that both $R_m^{\text{tabs}}$ and $R_m^{\text{air}}$ are inversely proportional to the cell spacing ($S_p$). Therefore, increasing $S_p$ reduces the CtC heat transfer.

\begin{figure}
  \centering
  \includegraphics[width = 1\textwidth]{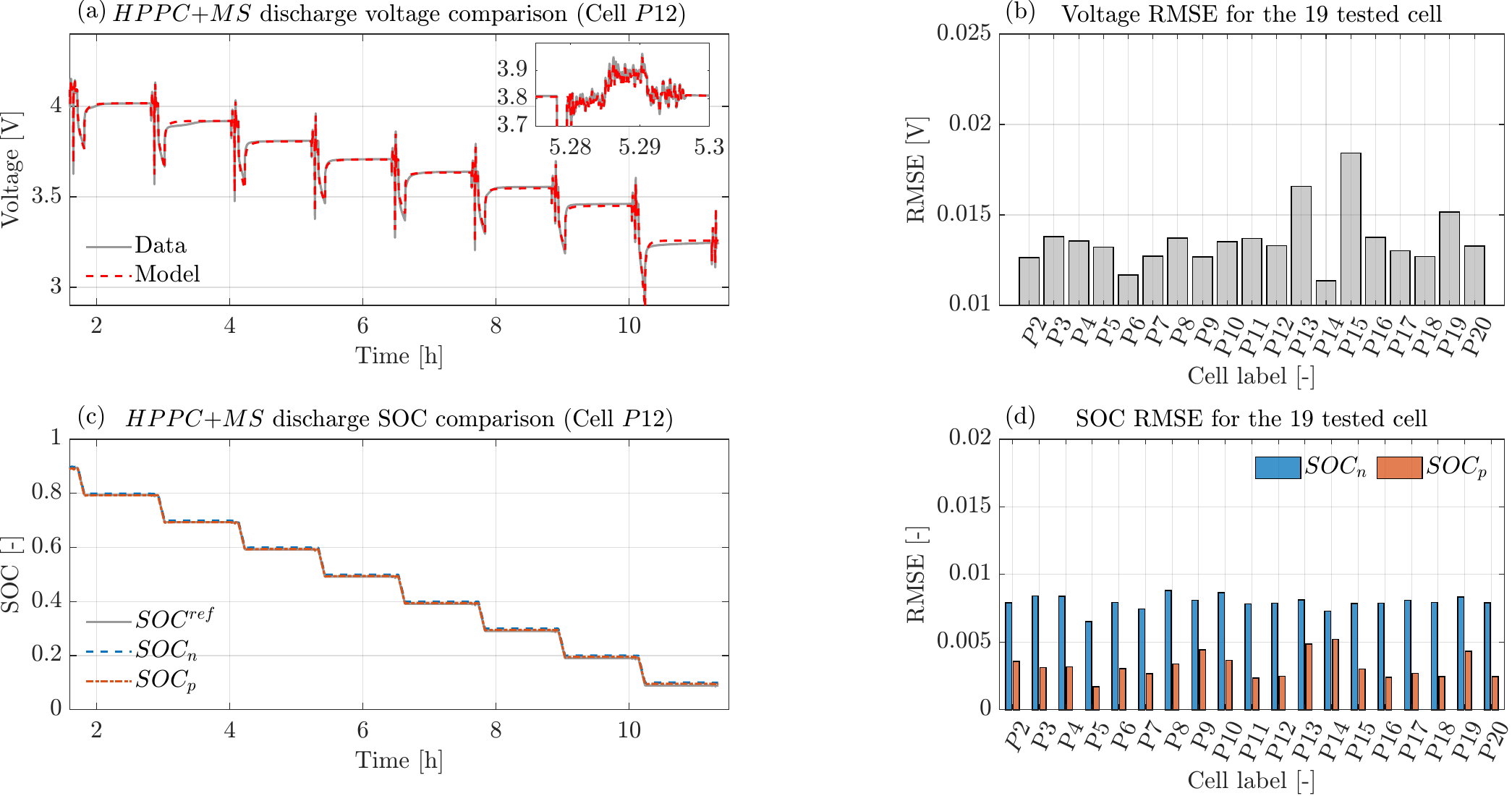}
  \caption{Single-cell parameter results for the 19 LG M50T cells. (a) and (c) voltage profile and
  SOC comparison between experiments and model simulations for cell P12 undergoing the validation cycle,
  respectively. (b) RMES between experimental voltage and model predictions for all cells. (d) RMSE between
  reference SOC and electrode SOC for all the cells.}
  \label{fig:Single_cell_identification}
\end{figure}

\subsection{Model identification and validation} \label{Sec:Model_validation}
In this section, we detail the identification and validation procedures for both cell-level, described in the \textbf{Supplementary Materials}, and module-level models. Specifically, the electrochemical parameters of the 19 LG M50T are determined using an optimization-based approach during the C/20 CC cycles. Subsequently, the resulting models are validated against each tested cell using the HPPC+MS cycle. Additionally, the identified ESPMs are combined to construct modules consisting of four cells connected in parallel. Both the thermal and cell current distributions within this module are then validated against experimental data. 

\subsubsection{Single-cell model identification and validation} \label{Sec:Cell_identification}
\begin{enumerate}
  \item \textbf{ESPMs identification}: The ESPM includes a total of 34 parameters\footnote{See Table S3 in the Supplementary Materials for the complete list}, spanning each cell domain and the electrolyte. 
  Although cell teardown analysis is currently the most advanced method for directly measuring a cell's physical, chemical, and electrochemical properties \cite{ecker2015parameterization}, this approach is complex, costly, and requires sophisticated equipment.
  Alternatively, optimization techniques can estimate model parameters by aligning simulated data with experimentally measured cell voltages \cite{bizeray2018identifiability}. However, due to the complexity and nonlinearity of the model, it is challenging to identify all parameters solely based on current-voltage measurements. Additionally, attempting to simultaneously identify all model parameters may result in overfitting, which can compromise the model's predictive accuracy and generalizability.
  
  In this work, we adopt a hybrid approach that combines both cell teardown and optimization-based techniques. Most of the ESPM parameters are sourced from \cite{chen2020development}, where the authors disassembled and measured the cell parameters of an LG M50, a previous version of the LG M50T. The exception is the active material volume fractions for both electrodes ($\epsilon_{s,j}$), which are individually identified for each of the 19 tested M50T cells using the C/20 constant current (CC) discharge cycle, following the procedure described in detail in \textbf{Supplementary Materials Note 4}.
  Overall, the performance of the 19 identified ESPMs is satisfactory, as demonstrated by the voltage RMSEs (Figure S1 of \textbf{Supplementary Materials}) ranging between 15 and 21 mV, and by the significant linear correlations between both $\epsilon_{s,j}$ and $Q_{cell}$, with $\epsilon_{s,n} = 0.0091055 + 0.16312 \cdot Q_{cell}$ and $\epsilon_{s,p} = 0.011719 + 0.14208 \cdot Q_{cell}$ exhibiting $R^2$ values of 0.993 and 0.967, respectively.
  
  \item \textbf{ESPMs validation}: The identified ESPMs are validated against the corresponding HPPC + MS cycle. Figures \ref{fig:Single_cell_identification}(a-d) summarize the model validation results. Specifically, Figures \ref{fig:Single_cell_identification}(a) and (c) offer a visual comparison between the measured and simulated voltages, and between the Coulomb counting SOC and the electrode SOC for cell P12, respectively. Overall, the ESPMs demonstrate a good fit for both cell voltage and SOC. This accuracy is underscored by the voltage RMSE for all tested cells (shown in Figure \ref{fig:Single_cell_identification}(b)), which ranges between 0.012 and 0.018 V. Additionally, the bar chart in Figure \ref{fig:Single_cell_identification}(d) displays the SOC RMSE for both electrodes across all cells, with a maximum RMSE of 0.9\% for the negative electrode and 0.51\% for the positive electrode.
\end{enumerate}

\begin{figure}
  \centering
  \includegraphics[trim=0 0 0 0,clip,width = 1\textwidth]{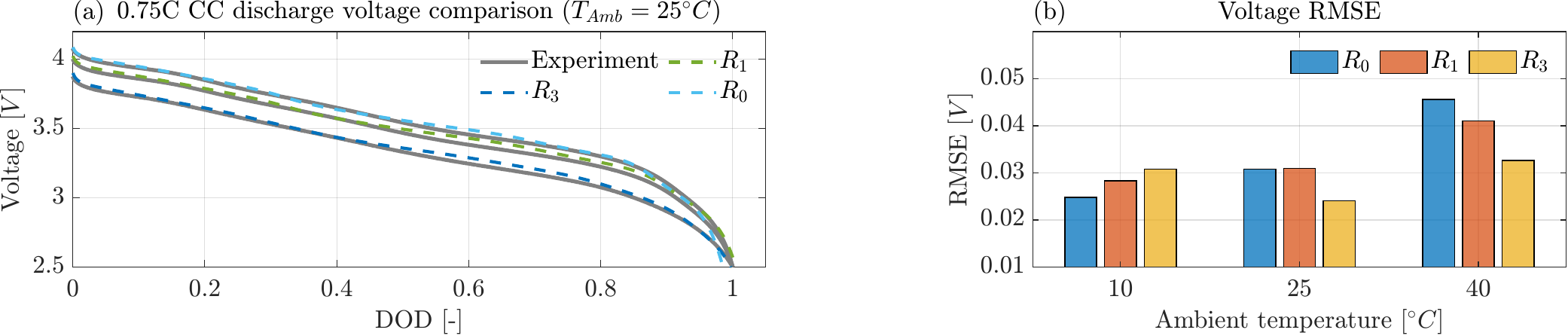}
  \caption{(a) Comparison of voltage profiles between experimental data and model simulations for $R_{int} = 0,1, \text{and } 3 \text{ m}\Omega$ at $25^{\circ} C$. (b) RMSE across all the considered experiment scenarios.}
  \label{fig:Module_RMSE}
\end{figure}

\subsubsection{Module-level model validation} \label{Sec:module_model_validation}
Following the parameterization of the cell-level dynamics, this section addresses the next critical step: verifying the module model ability to accurately predict both electrical and thermal dynamics within the module.
In this study, we evaluate nine distinct configurations of a module composed of four cells connected in parallel. These configurations are systematically varied based on combinations of interconnection resistance ($R_{int} =$ 0, 1, and 3 $m\Omega$) and ambient temperature ($T_{amb} =$ 10$^{\circ} C$, 25$^{\circ} C$, and 40$^{\circ} C$).

\begin{figure}
  \centering
  \includegraphics[trim=0 0 0 0,clip,width = 1\textwidth]{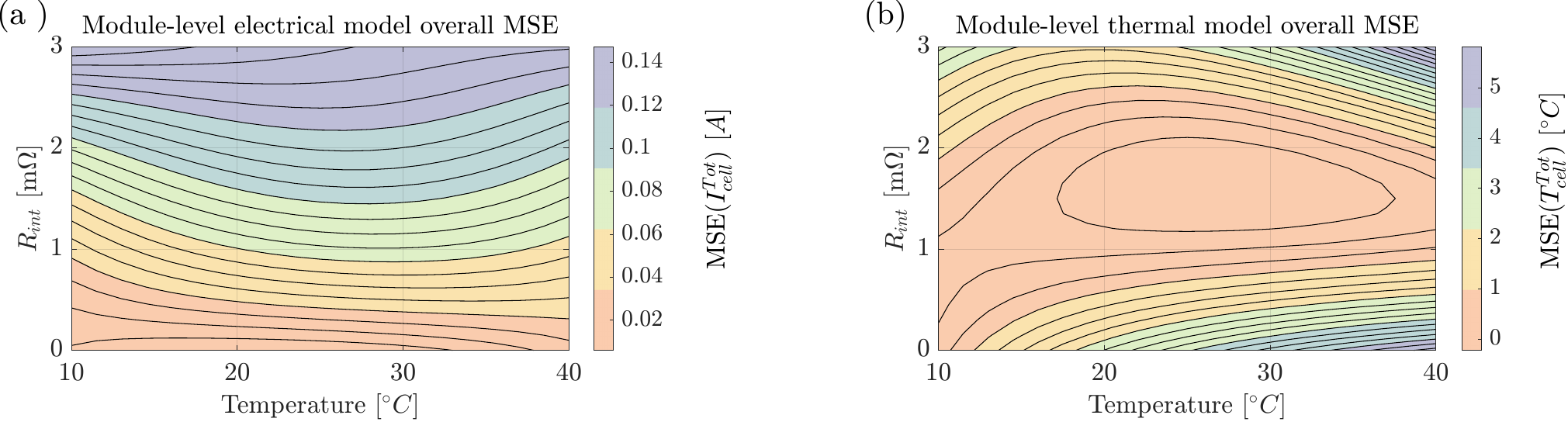}
  \caption{Contour plots of the cumulative MSE for (a) cell currents and (b) temperatures. The cumulative MSE for cell currents, $\sum_{k=1}^{N_p} \text{MSE}(I_{\text{cell}}^{[k]})$, and for temperatures, $\sum_{k=1}^{N_p} \text{MSE}(T_{\text{cell}}^{[k]})$, are calculated based on the MSE values reported in Table \ref{Tab:MSE}.
  \label{fig:contour_plots}}
\end{figure}

The verification process involves comparing the model predictions of module overall voltage, cell currents, and temperatures of each cell against experimental data for each configuration throughout the entire discharge cycle. Particularly, the main trends are:
\begin{itemize}
\item \textbf{Module voltage:} Figure \ref{fig:Module_RMSE} (a) compares the measured and simulated voltages for $R_{int} = 0,1, \text{ and } 3 \text{ m}\Omega$ at ambient temperature of $25 ^{\circ} C$. Additionally, Figure \ref{fig:Module_RMSE} (b) presents a bar chart illustrating the voltage RMSE across all module experimental scenarios.
While the overall model performance shows a slight deterioration compared to the single-cell results, particularly at high temperatures and low interconnection resistances, it remains within acceptable limits, with a maximum voltage RMSE of $45.6$ mV observed across all scenarios.

\item \textbf{Cell currents:} The model fitting accuracy was assessed by calculating the MSE between the cell currents estimated by the module and those measured by the Hall sensor ($MSE(I_{cell}^{[k]})$), as defined in Equation \eqref{eq:MSE_Icell}. The MSE values for each cell within the nine module configurations are listed in Table \ref{Tab:MSE}. 
Notably, during a complete CC discharge cycle at a 0.75 C-rate, the error remains below 77.8 $mA$ across all 36 cells considered, which represents approximately 2\% of the reference cell current value at 0.75C-rate (i.e., 0.75$\times$4.85 $Ah$).
Additionally, the cumulative error for each module ($\text{MSE}(I_{\text{cell}}^{Tot}$)\footnote{$\text{MSE}(I_{\text{cell}}^{Tot}) = \frac{1}{N\cdot N_p}\sum_{k=1}^{N_p} \sum_{i=1}^{N} (I_{cell}^{[k]}|_{t=i}^{sim} - I_{cell}^{[k]}|_{t=i}^{data})^2$.})  is visualized in the contour plots presented in Figure \ref{fig:contour_plots}(a), illustrating that the current error increases with rising interconnection resistance while remaining insensitive to variations in ambient temperature.

\item \textbf{Cell temperatures:} Note that, the thermal dynamics of the cells are identified at the module level considering a CC discharge at 0.75C-rate, considering the scenario with $R_{int} =$ 1 $\text{ m}\Omega$ at 25$^{\circ} C$ following the procedure described in  \textbf{Supplementary Materials Note 5}.

The thermal model is  validated across all nine module configurations by calculating the MSE between the simulated and measured cell surface temperatures for each cell within the module ($MSE(T_{\text{cell}^{[k]}})$), as presented in Table \ref{Tab:MSE}. Additionally, the cumulative MSE per module ($\text{MSE}(T_{\text{cell}}^{Tot}$)\footnote{$\text{MSE}(T_{\text{cell}}^{Tot}) = \frac{1}{N\cdot N_p}\sum_{k=1}^{N_p} \sum_{i=1}^{N} (T_{cell}^{[k]}|_{t=i}^{sim} - T_{cell}^{[k]}|_{t=i}^{data})^2$.})  is visualized in the contour plots shown in Figure \ref{fig:contour_plots}(b). These visualizations indicate that the model performs satisfactorily under conditions of low interconnection resistance ($R_{int}$) and ambient temperature, with a maximum cumulative MSE of 2 $^{\circ}$C. However, the model's performance declines at an ambient temperature of $40^{\circ}$C.
Overall, the validation of the thermal model is influenced by the thermal sensor tolerances (i.e., $\pm$ 0.5 $^{\circ}$C), especially when thermal gradients are minimal. Furthermore, parasitic resistance introduced during module assembly, which is not fully accounted for in the model formulation, contributes to the reduced accuracy. Future work will explore the possibility of enhancing model accuracy by developing more complex models.
\end{itemize}
It is important to note that during the validation process, the electrochemical parameters for each cell within the model remain consistent with those identified at the cell level in Section \ref{Sec:Cell_identification}. Therefore, this approach not only tests the model capability to predict overall module performance and heterogeneities under different scenarios but also demonstrates that the model can predict module imbalances by scaling up CtC variations to the module level.

\begin{table*}
	\renewcommand{\arraystretch}{2}
	\caption{MSEs between the simulated and measured cell currents and temperatures for each tested module.}
	\scriptsize
	\label{Tab:MSE}
	\centering
  \resizebox{1\textwidth}{!}{	
    \begin{tabular}{m{1cm} | m{1cm} | m{1cm} m{1cm} m{1cm} m{1cm} | m{1cm} m{1cm} m{1cm} m{1cm}}
    \hline\hline 
    \multicolumn{10}{c}{\parbox{0.85\textwidth}{ \begin{equation} \label{eq:MSE_Icell} 
      \text{MSE}(I_{cell}^{[k]}) = \frac{1}{N} \sum_{i=1}^{N} (I_{cell}^{[k]}|_{t=i}^{sim} - I_{cell}^{[k]}|_{t=i}^{data})^2
    \end{equation}}} \\
    \multicolumn{10}{c}{\parbox{0.85\textwidth}{ \begin{equation} \label{eq:MSE_Tcell} 
      \text{MSE}(T_{cell}^{[k]}) = \frac{1}{N} \sum_{i=1}^{N} (T_{cell}^{[k]}|_{t=i}^{sim} - T_{cell}^{[k]}|_{t=i}^{data})^2
    \end{equation}}} \\
    \hline
    \hfil \multirow{2}{*}{$T_{amb}$} & \hfil \multirow{2}{*}{$R_{int}$} & \multicolumn{4}{c}{\textit{Cell current MSE [A]}} & \multicolumn{4}{c}{\textit{Cell temperature MSE [ $^{\circ} C$]}}  \\ 
    \cline{3-6} \cline{7-10}
    & & Cell 1 & Cell 2 & Cell 3 & Cell 4 & Cell 1 & Cell 2 & Cell 3 & Cell 4  \\
    \hline 
    \hfil \multirow{3}{*}{10 $^{\circ} C$} & \hfil 0 $m \Omega$ & 0.0047 & 0.0029 & 0.0013 & 0.0035 & 0.1445 & 0.2951  &  0.0432  &  0.0488
    \\ [-0.5em]
    & \hfil 1 $m \Omega$ & 0.0201  &  0.0047  &  0.0079  &  0.0046 & 0.1562 & 0.0441  &  0.0924  &  0.0423 \\  [-0.5em]
    & \hfil 3 $m \Omega$ & 0.0778  &  0.0192  &  0.0185  &  0.0388 & 0.6381 & 1.0267  &  0.4589  &  0.7598 \\ [-0.25em]
    \hline
    \hfil \multirow{3}{*}{25  $^{\circ} C$} & \hfil 0 $m \Omega$ & 0.0013  &  0.0014  &  0.0032  &  0.0008 & 0.8457 &   2.0558  &  0.1871 &   0.2693 \\ [-0.5em]
    & \hfil 1 $m \Omega$ & 0.0214  &  0.0026  &  0.0090  &  0.0349  & 0.1250   & 0.0418   & 0.0565  &  0.1102 \\  [-0.5em]
    & \hfil 3 $m \Omega$ & 0.0669  &  0.0218  &  0.0208  &  0.0320  & 0.4028  &  0.9291  &  0.3519  &  0.5703 \\ [-0.25em]
    \hline
    \hfil \multirow{3}{*}{40 $^{\circ} C$} & \hfil 0 $m \Omega$ & 0.0061  &  0.0054  &  0.0069  &  0.0055 & 1.0453   & 2.0555  &  0.7285   & 1.5308 \\ [-0.5em]
    & \hfil 1 $m \Omega$ & 0.0310  &  0.0031  &  0.0054  &  0.0185 & 0.3151  &  0.0994  &  0.2536   & 0.0464 \\  [-0.5em]
    & \hfil 3 $m \Omega$ & 0.0674  &  0.0186  &  0.0209  &  0.0271 & 1.2478  &  2.1062  &  1.3439  &  1.4218 \\ 
		\hline\hline 
	\end{tabular}}
\end{table*}

\subsection{Multi-linear regression model} \label{Sec:MLR}
The statistical analysis is conducted using a technique called multi-linear regression (MLR). MLR is a statistical method employed to determine the influence of a set of independent variables (predictors) on a response variable, assuming a linear relationship between the inputs and the output.
Consider a set of $N$ observations, denoted as $\Theta = \{(X_1, y_1), (X_2, y_2), \cdots, (X_N, y_N)\}$, where $X_k = [x_1, \cdots, x_q] \in \mathbb{R}^q$ and $y_k \in \mathbb{R}$ (with $k = 1, \cdots, N$) represent the vector containing the $q$ predictors ($x_j, j = 1, \cdots, q$) and the response variable for the $k$-th sample, respectively.
According to \cite{hastie2009elements}, the MLR model can be formulated as follows:
\begin{equation} \label{eq:MLR}
  \hat{y} = f(x_1, x_2, \cdots, x_q) = \beta_0 + \sum_{z = 1}^{q} (\beta_z x_z + \beta_{zz} x_z^2)  + \sum_{z = 1}^{q-1} \sum_{y = i+1}^{q} (\beta_{z,y} x_z x_y)
\end{equation}
where $\beta_0$ represents the intercept, $\beta_z$ and $\beta_{zz}$ are constant coefficients associated with the linear and quadratic terms, while $\beta_{z,y}$ corresponds to the interactions between variables. It is worth noting that the MLR model can also include nonlinear terms, such as polynomials or interactions between variables, as they can be considered equivalent to additional predictors impacting the response variable linearly.
The coefficients of the MLR model in eq. \eqref{eq:MLR} are estimated using a least square approach that minimizes the sum of squared residuals:
\begin{equation}
  \min_{\beta_0, \beta_z, \beta_{zz}, \beta_{z,y}}  \sum_{k=1}^{N} \epsilon_k =  \min_{\beta_0, \beta_z, \beta_{zz}, \beta_{z,y}}  \sum_{k=1}^{N} (y_k - \beta_0 - \sum_{z = 1}^{q} (\beta_z x_z + \beta_{zz} x_z^2)  - \sum_{z = 1}^{q-1} \sum_{y = i+1}^{q} (\beta_{z,y} x_z x_y))
\end{equation}
Since not all terms in the MLR model may be statistically significant, their relevance is examined by testing the null hypothesis (NH) for each coefficient using p-value analysis \cite{thiese2016p}. For example, considering the coefficient $\beta_z$, the NH assumes that there is no relationship between $y$ and $x_z$, meaning $\beta_z = 0$. The p-value represents the probability of observing the NH, and it is calculated based on the t-statistic of $\beta_z$ \cite{hastie2009elements}. A high p-value indicates that the NH is likely true, suggesting that the term associated with $\beta_z$ is not statistically significant and can be disregarded.
In this study, a reduced-order model (ROM) of Eq. \eqref{eq:MLR} is derived by considering only the significant terms with a p-value greater than 0.05 \cite{thiese2016p}. The ROM formulation is a crucial step in the statistical analysis as it enables the identification of the most relevant predictors, thereby highlighting potential significant nonlinear relationships between the inputs and outputs, as well as important interactions among the predictors.
To achieve this, we utilized stepwise regression for automated predictor selection \cite{hocking1976biometrics}.
Stepwise regression \footnote{Stepwise regression is performed employing MatLab \texttt{stepwiselm} function.} systematically identifies the model that includes only statistically significant predictors, based on specific criteria (p-values in our case), employing both forward and backward feature selection methods.
To assess the accuracy of the MLR model, the coefficient of variation ($R^2$) is utilized:
\begin{equation}
  R^2 = 1 - \frac{\sum_{k=1}^{N} (y_k-\hat{y}_k)^2}{\sum_{k=1}^{N} (y_k-\frac{1}{N} \sum_{k=1}^{N} y_k)^2}
\end{equation}
Here, $R^2$ ranges between 0 and 1, indicating the proportion of variability in the response variable captured by the MLR model. A value of 1 signifies that the model precisely captures the variability, while lower values indicate a lesser degree of captured variability.
One limitation of this methodology is its inflexibility, as it relies on a predefined structure for the regression model, which may introduce potential inaccuracies.

\begin{figure}
  \centering
  \includegraphics[trim=0 0 0 0,clip,width = 0.9\textwidth]{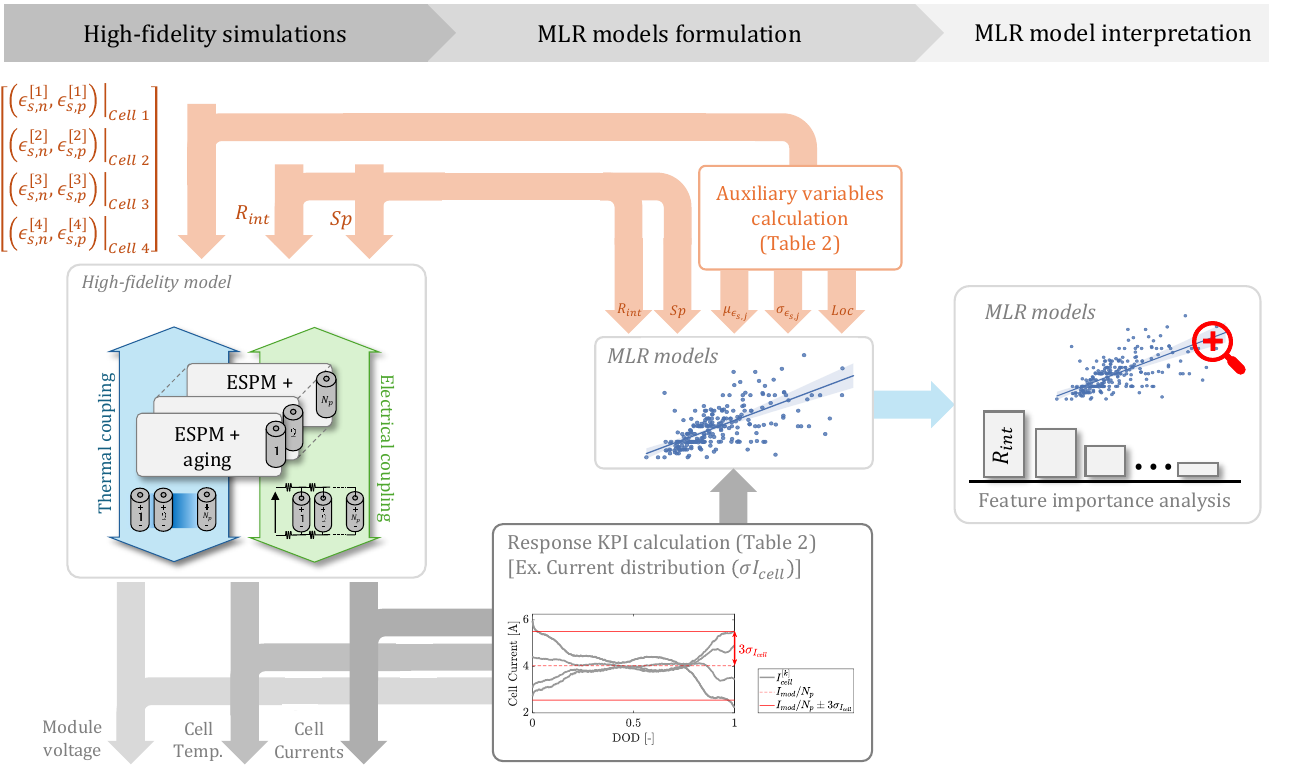}
  \caption{Flowchart illustrating the process used in this study to analyze the effects of CtC variations on parallel-connected modules, as described in Section~\ref{Sec:Res_and_disc}. The figure highlights the response variable $\sigma I_{cell}$ as an example, while other response variables are listed in Table~\ref{Tab:MLR}.}
  \label{fig:MLR_overview}
\end{figure}

\section{Results and discussion} \label{Sec:Res_and_disc}
Leveraging the previously validated module-level model, in this section we shown the MLR analysis results applied to the paralle-connected battery module scenario.
The objective of the analysis proposed in this paper is to identify the key battery cell and module parameters that exert the most significant influence on the variations in cell current and temperature distributions, as well as, on the energy and capacity of the module under both fresh and aged conditions (evaluated considering the cell Solid Electrolyte Interface (SEI) layer growth).

Particularly, the analysis is performed follwing three steps as visually represented in Figure \ref{fig:MLR_overview}:
\begin{enumerate}
  \item \textbf{Hihg-fidelity offline simulations: } A total of 500 battery modules are generated, each of them undergoes 500 cycles at an ambient temperature of $T_{amb} = 25^{\circ} C$. Each cycle consists of the following steps: 1) CCCV charge at a C/3-rate, 2) 30 minutes of rest, 3) CC discharge at a 1C-rate, starting from 100\% SOC down to 0\% SOC, 4) another 30 minutes of rest to allow for the balancing of any SOC heterogeneity within the module.
  
  Each battery module simulation incorporates common sources of CtC variations. Specifically, the configuration of each module is randomly determined by sampling values of the electrode active material volume fractions \(\epsilon_{s,n}\) and \(\epsilon_{s,p}\) per each cell, electrical interconnection resistances (\(R_{\text{int}}\)), as well as cell spacing (\(Sp\)) and location. Variations in \(R_{\text{int}}\) and \(Sp\) arise from differences in module design and manufacturing precision. Additionally, perturbation of \(\epsilon_{s,n}\) and \(\epsilon_{s,p}\) are introduced to alter the capacity of each individual cell, resulting from tolerances in various electrode production steps \cite{niri2022systematic}. Note that, the \(\epsilon_{s,n}\) and \(\epsilon_{s,p}\) are simultaneously perturbed for each cell in the module.

  It is worth highlighting, that the rationale behind selecting \(\epsilon_{s,n}\) and \(\epsilon_{s,p}\), to induce capacity heterogeneities among single cell, over the others parameters in the ESPM is driven by two main reasons.
  Firstly, according to \cite{andersson2022parametrization}, electrode capacity-related parameters such as the electrode active material volume fraction (\(\epsilon_{s,j}\)) and thickness (\(L_j\)) consistently exhibit the highest sensitivity under both constant current and driving cycle scenarios. Perturbing these parameters directly impacts the single cell capacity.
  Further, in the ESPM identification process, Section \ref{Sec:Cell_identification}, we demonstrate that variations in \(\epsilon_{s,j}\) effectively capture capacity heterogeneities in batches of fresh cells. 

  \item \textbf{MLR Model Formulation:}  
  Given the simulated dataset, the MLR models are constructed to analyze the relationships between predictors and response variables. The predictors for each MLR model are selected based on the CtC sources introduced during the offline simulations. As schematically shown in Figure \ref{fig:MLR_overview}, the electrical interconnection resistances and cell spacing values are directly used as model features. In contrast, the heterogeneities in the electrode active material volume fractions (\(\epsilon_{s,n}\) and \(\epsilon_{s,p}\)) for each cell are represented using auxiliary variables that characterize the parameter distribution within the module, as detailed in Section \ref{Sec:MLR_module}. These auxiliary variables are more appropriate for assessing sensitivity because the focus is on the effect of the parameter distribution within the module rather than on individual cell parameters. 
  
  Furthermore, the MLR response variables are calculated based on voltage, temperature, and cell current data resulting from the high-fidelity simulations. Key outputs, such as module capacity, energy, current, thermal profiles, and aging heterogeneities, are evaluated as clarified in the next Section.

  \item \textbf{MLR model interpretation:} Given the Forumalted MLR models,  a relative importance analysis is performed, as outlined in \cite{tonidandel2011relative} and detailed in Section \ref{Sec:MLR}, to rank the predictors based on their influence on the module's desired response variables.
\end{enumerate}

\begin{table}
	\caption{Multi linear regression model predictor and response variable}
	\footnotesize
	\label{Tab:MLR}
	\centering
  \resizebox{1\textwidth}{!}{	
    \begin{tabular}{m{1.5cm} m{1cm} m{1cm} m{1cm} m{1cm} m{1cm} m{1cm} m{1cm} m{1cm} m{1cm} | m{1.5cm}}
    \hline\hline \\[-1em]
    \multicolumn{11}{l}{\textbf{MLR models: Predictors}} \\ [-0.2 em]
    \hline \\ [-2em]
    \multicolumn{5}{c}{\parbox{0.48\textwidth}{ \begin{equation} \label{eq:AVG} 
      \mu_{\epsilon_{j}} = \frac{1}{N_p} \sum_{i=1}^{N_p} \epsilon_{s, j}^{[i]},  \quad \text{with: } j \in [n,p]
     \end{equation}}} &
   \multicolumn{6}{c}{\parbox{0.48\textwidth}{ \begin{equation} \label{eq:RANGE} 
      \sigma_{\epsilon_{j}} = \sqrt{ \frac{1}{N_p-1} \sum_{z=1}^{N_p} (\epsilon_{s, j}^{[z]} - \mu_{\epsilon_{j}})^2},  \quad j \in [n,p]
   \end{equation}}}\\ [-0em]

    \multicolumn{11}{l}{\parbox{1\textwidth}{ \begin{equation} \label{eq:Loc_index} 
      Loc = \sum_{i=1}^{N_p} w_i \cdot \text{min}(\bar{\epsilon_{s,n}}^{[i]}, \bar{\epsilon_{s,p}}^{[i]}), \qquad \quad \text{where} \qquad 
      w_i = \begin{cases} 
      (\frac{N_p}{2}+1) - P_i \quad  &\text{if: } 1 \leq P_i \leq \frac{N_p}{2} \\
       \frac{N_p}{2} - P_1 \qquad &\text{if: } \frac{N_p}{2} < P_i \leq N_p
      \end{cases}
    \end{equation}}} \\ [-0.4em]

    \multicolumn{5}{c}{\parbox{0.48\textwidth}{ \begin{equation} \label{eq:mu_comb} 
      \mu_{comb} = \frac{1}{N_p} \sum_{i=1}^{N_p} min(\bar{\epsilon_{s, n}}^{[i]}, \bar{\epsilon_{s, p}}^{[i]})
     \end{equation}}} &
   \multicolumn{6}{c}{\parbox{0.48\textwidth}{ \begin{equation} \label{eq:sigma_comb} 
      \sigma_{comb} = \sqrt{ \frac{1}{N_p-1} \sum_{z=1}^{N_p} (min(\bar{\epsilon_{s, n}}^{[i]}, \bar{\epsilon_{s, p}}^{[i]}) - \mu_{comb})^2}
   \end{equation}}}\\ [-0em]
        
   \multicolumn{5}{c}{\parbox{0.48\textwidth}{ \begin{equation} \label{eq:Loc_j} 
    Loc_j = \sum_{i=1}^{N_p} w_i \cdot \text{min}(\epsilon_{s,j}^{[i]}),  \quad j \in [n,p]
   \end{equation}}} & \\ [-0em]

 \hline\hline \\[-1em]
 \multicolumn{11}{l}{\textbf{MLR models: Response variables}} \\ 
 \hline \\ [-1em]

 \multicolumn{11}{c}{\parbox{0.98\textwidth}{ \begin{equation} \label{eq:sigma_Icell}  
  \sigma I_{cell}=  \frac{1}{t_{end}-t_0} \int_{t_0}^{t_{end}} \sqrt{ \frac{1}{N_p-1} \sum_{z=1}^{N_p} (I_{cell}^{[z]} - I_{mod}/N_p)^2} dt
\end{equation}}} \\ 

\multicolumn{11}{c}{\parbox{0.98\textwidth}{ \begin{equation}  \label{eq:sigma_Tcell}
  \sigma T_{cell}=  \frac{1}{t_{end}-t_0} \int_{t_0}^{t_{end}}  \sqrt{ \frac{1}{N_p-1} \sum_{z=1}^{N_p} \Big( T_{cell}^{[z]} - \frac{1}{N_p} \sum_{i=1}^{N_p} T_{cell}^{[i]}\Big)} dt
\end{equation}}} \\ 

    \multicolumn{5}{c}{\parbox{0.48\textwidth}{ \begin{equation} \label{eq:DE} 
      \%\Delta E = 100\frac{E_{mod}-E_{mod}^{ref}}{E_{mod}^{ref}}
      \end{equation}}} &
    \multicolumn{6}{c}{\parbox{0.48\textwidth}{ \begin{equation} \label{eq:DC} 
        \%\Delta Q = 100\frac{Q_{mod}-Q_{mod}^{ref}}{E_{mod}^{ref}}
        \end{equation}}}\\ [-0em]

    \multicolumn{5}{c}{\parbox{0.48\textwidth}{ \begin{equation}  \label{eq:DT} 
      \Delta T_{max} = max(T_{s}^{mod}) - min(T_{s}^{mod})
      \end{equation}}} &
    \multicolumn{6}{c}{\parbox{0.48\textwidth}{ \begin{equation} \label{eq:Elost}  
      E_{lost} = E_{mod}^{EOS}-E_{mod}^{BOL}
      \end{equation}}}\\ 

      \multicolumn{11}{c}{\parbox{0.98\textwidth}{ \begin{equation}  \label{eq:sigma_Rsei}
        \sigma R_{SEI}^{EOS}= \sqrt{ \frac{1}{N_p-1} \sum_{z=1}^{N_p} \Big( R_{SEI}^{[z], EOS} - \frac{1}{N_p} \sum_{i=1}^{N_p} R_{SEI}^{[i]} \Big)^2}
        \end{equation}}} \\ 
		\hline\hline \\[-8mm]
	\end{tabular}}
\end{table}

\subsection{CtC variation effect: MLR analysis} \label{Sec:MLR_module}
A total of 7 MLR models are considered in this work. Note that all MLR models share the same set of predictors, while each model has a unique response variable. Specifically:

\paragraph{MLR Predictors:}

The  electrical interconnection resistances and cell spacing values are directly used as predictor in the MLR model. Particularly, the cell spacing is assumed to be identical across all cells, and is randomly sampled within the interval \([1, 10]\) mm for each module, while the  (\(R_{\text{int}}\)) is randomly sampled within the interval \([0.1, 0.5]\) m\(\Omega\).

As mentioned earlier, the (\(\epsilon_n\) and \(\epsilon_p\)) are simultaneously perturbed for each cell in the module, and auxiliary variables, specifically the mean value (\(\mu_{\epsilon_{s,j}}\)) and standard deviation (\(\sigma_{\epsilon_{s,j}}\)) of both parameters, are considered as predictors to account for the parameter distribution within the module. These are calculated as follows:
    \begin{equation} 
      \mu_{\epsilon_{s,j}} = \frac{1}{N_p} \sum_{i=1}^{N_p} \epsilon_{s,j}^{[i]}, \qquad \sigma_{\epsilon_{s,j}} = \sqrt{\frac{1}{N_p-1} \sum_{i=1}^{N_p} \left(\epsilon_{s,j}^{[i]} - \mu_{\epsilon_{s,j}}\right)^2}, \qquad \text{for } j \in \{n, p\}
    \end{equation}
It is important to note that, assuming cell capacity can deviate by up to 2.5\% from the nominal value (i.e., 4.85 Ah for the LG M50T battery), the lower and upper bounds for \(\epsilon_n\) and \(\epsilon_p\) are determined based on the relationships identified in Section \ref{Sec:Cell_identification}, considering the batch of 19 tested cells. Specifically, the values of \(\epsilon_n\) and \(\epsilon_p\) are randomly sampled for each cell, assuming a uniform probability distribution for both parameters.
Finally, a location index is considered as additional predictor. The introduction of the predictor \(Loc\) aims to consider the effect of cell arrangement on the thermal gradient, and is calcualted as detailed in Table \ref{Tab:MLR}. Specifically, for each perturbed parameter, \(Loc\) is calculated as a weighted mean value of \(\min(\bar{\epsilon_{s,n}}^{[i]}, \bar{\epsilon_{s,p}}^{[i]})\), as shown in Equation \eqref{eq:Loc_index}. Here, \(\bar{\epsilon_{s,j}}^{[i]}\) is the normalized value of \(\epsilon_{s,j}^{[i]}\) within the interval [0,1], and \(w_i\) is the weight depending on the position of the cell (\(P_i =1,2,\dots, N_p\)) within the module.

\paragraph{MLR response variables:}

The quantities of interest for the CtC analysis (i.e. response variables) that are computed for all the conducted simulations are presented in Table \ref{Tab:MLR}. 
In particular, $\%\Delta E$ and $\%\Delta Q$, calculated as in \eqref{eq:DE} and \eqref{eq:DC}, indicate the percentage deviation of the module energy ($E_{mod}$) and capacity ($Q_{mod}$) from their reference values $E_{mod}^{ref}$ and $Q_{mod}^{ref}$, respectively.
The reference module is composed by 4 unperturbed cells, cycled considering $R_{int} = 0.25 m\Omega$ and $Sp = 5 mm$, where $E_{mod}$ and $Q_{mod}$  are calculated as: 
\begin{equation} \label{eq:Emod_Qmod}
  E_{mod} = \int_{t_{in}^{ch}}^{t_{end}^{ch}} V_{mod}I_{mod} dt,  \qquad \qquad Q_{mod} = \int_{t_{in}^{ch}}^{t_{end}^{ch}} I_{mod} dt
\end{equation}
and $t_{in}^{ch}$ and $t_{end}^{ch}$ are the initial and final time instant of the CC discharging cycle.
Further, $\sigma_{I_{cell}}$ is the mean of the cell current standard deviation, calculated as in \eqref{eq:sigma_Icell}. 
$\Delta T_{max}$ is maximum thermal gradient \eqref{eq:DT} and $\sigma_{T_{cell}}$ is the mean of the cell temperature standard deviation.
$E_{lost}$  measures the energy lost experienced from the first charging cycle (i.e., the Beginning Of Simulation - BOS) until the End of the Simulation (EOS), after a total of 500 cycles.
$\sigma R_{SEI}^{EOS}$, reported in Eq. \eqref{eq:sigma_Rsei}, represents the standard deviation of SEI resistance of the cells within the module at the EOS.

\begin{figure}
	\centering
	\includegraphics[trim=0 0 0 0,clip,width = 1\textwidth]{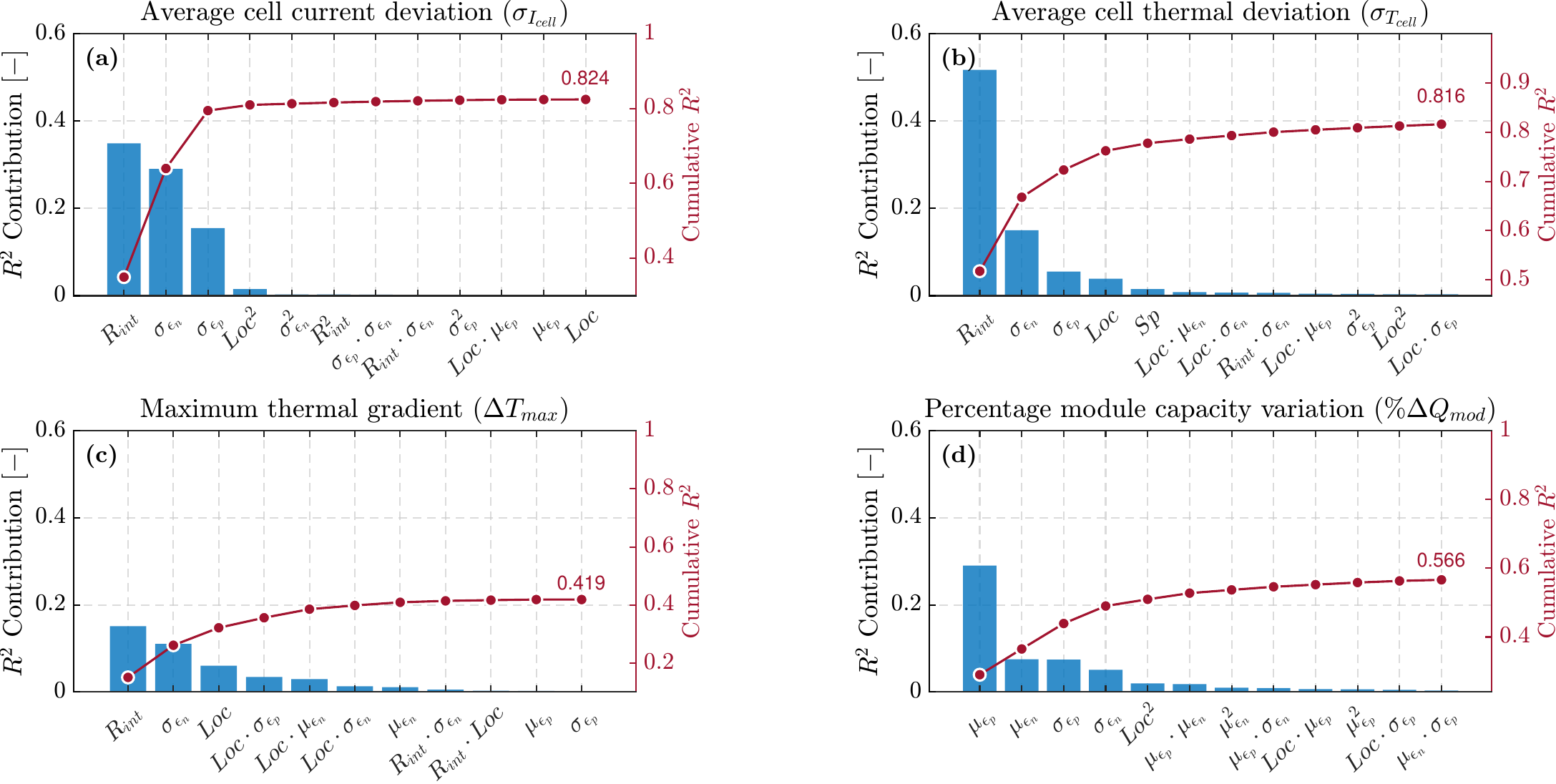}
  \includegraphics[trim=0 0 0 0,clip,width = 0.5\textwidth]{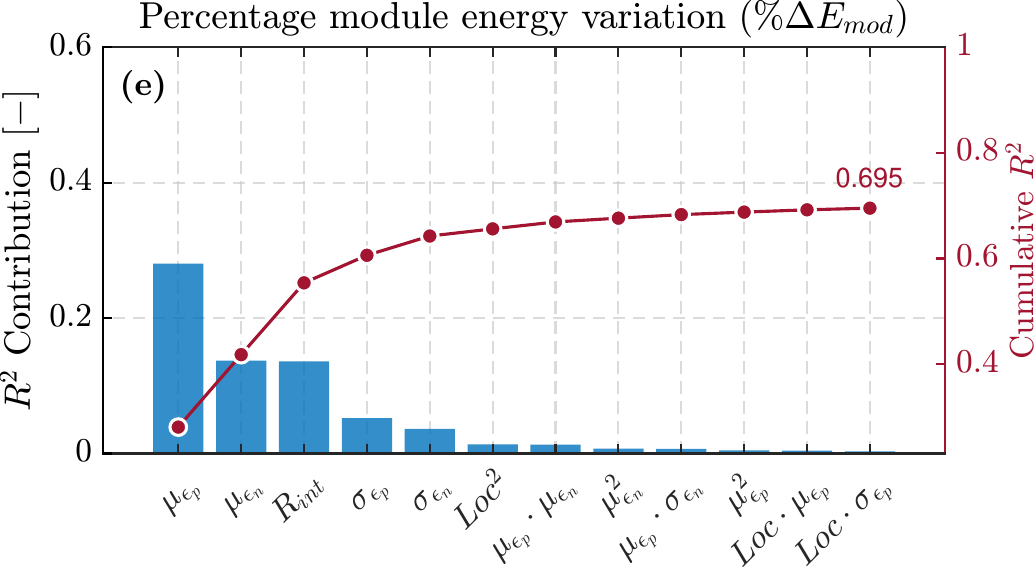}
  \caption{Figures (a-e) depict the Pareto plot for each considered response variable in the short-term scenario. Each subplot shows the contribution of predictors to the $R^2$ of the MLR model via bar charts, accompanied by a solid line representing the cumulative $R^2$.}
  \label{fig:Pareto_1}
\end{figure}

\subsubsection{MLR analysis: short term CtC heterogeneity effects} \label{Sec:shortTerm_analysis}
In this section, the short-term impacts of CtC heterogeneities on fresh battery cells are investigated. 
We focus on the previously mentioned response variables, specifically $\%\Delta Q$, $\%\Delta E$, $\sigma_{I_{cell}}$, $\sigma_{T_{cell}}$ and $\Delta T_{max}$, while excluding those affected by aging, such as $E_{lost}$, and $\sigma R_{SEI}^{EOS}$. 
The MLR models are then constructed using the structure defined in Equation \eqref{eq:MLR}, focusing on terms that exhibit statistical significance (p-value $<$ 0.05). 
To evaluate the relative importance of each predictor, a comprehensive relative importance analysis was performed, and the results are presented in Figure \ref{fig:Pareto_1} for each considered response variable.
This analysis assesses the contribution of each predictor to the increase in the corresponding model's $R^2$ value. 

Additionally, to enhance the explanation of the variability in the module response and to capture the key factors influencing the module's behavior, an additional seven predictors are added to the MLR model depending on the response variable considered. These additional predictors are: $\mu_{comb}$, $\sigma_{comb}$, $Loc_n$, $Loc_p$, $\sigma SOC$, $\mu SOC$, and $\Delta T_{max}$.
$\mu_{comb}$ and $\sigma_{comb}$ represent the lumped mean and standard deviation of $\epsilon_{s,j}$, accounting for the fact that cell capacity is limited by the weakest electrode. The $\epsilon_{s,j}$ values are normalized to the [0,1] range, and the average is calculated by taking the minimum $\epsilon_{s,j}$ value for each cell, as detailed in equation \eqref{eq:mu_comb}.
$Loc_j = \sum_{i=1}^{N_p} w_i\epsilon_{s,j}^{[i]}$ with $j \in [n,p]$ stands for  the location index for each perturbed cell-level parameter.
$\sigma SOC$ and $\mu SOC$ are the standard deviation and mean value of the SOC of the $N_p$ cells at the end of the discharge cycle.
The main trends resulting form the MLR analysis are the following:

\begin{figure}
	\centering
	\includegraphics[trim=0 0 0 0,clip,width = 1\textwidth]{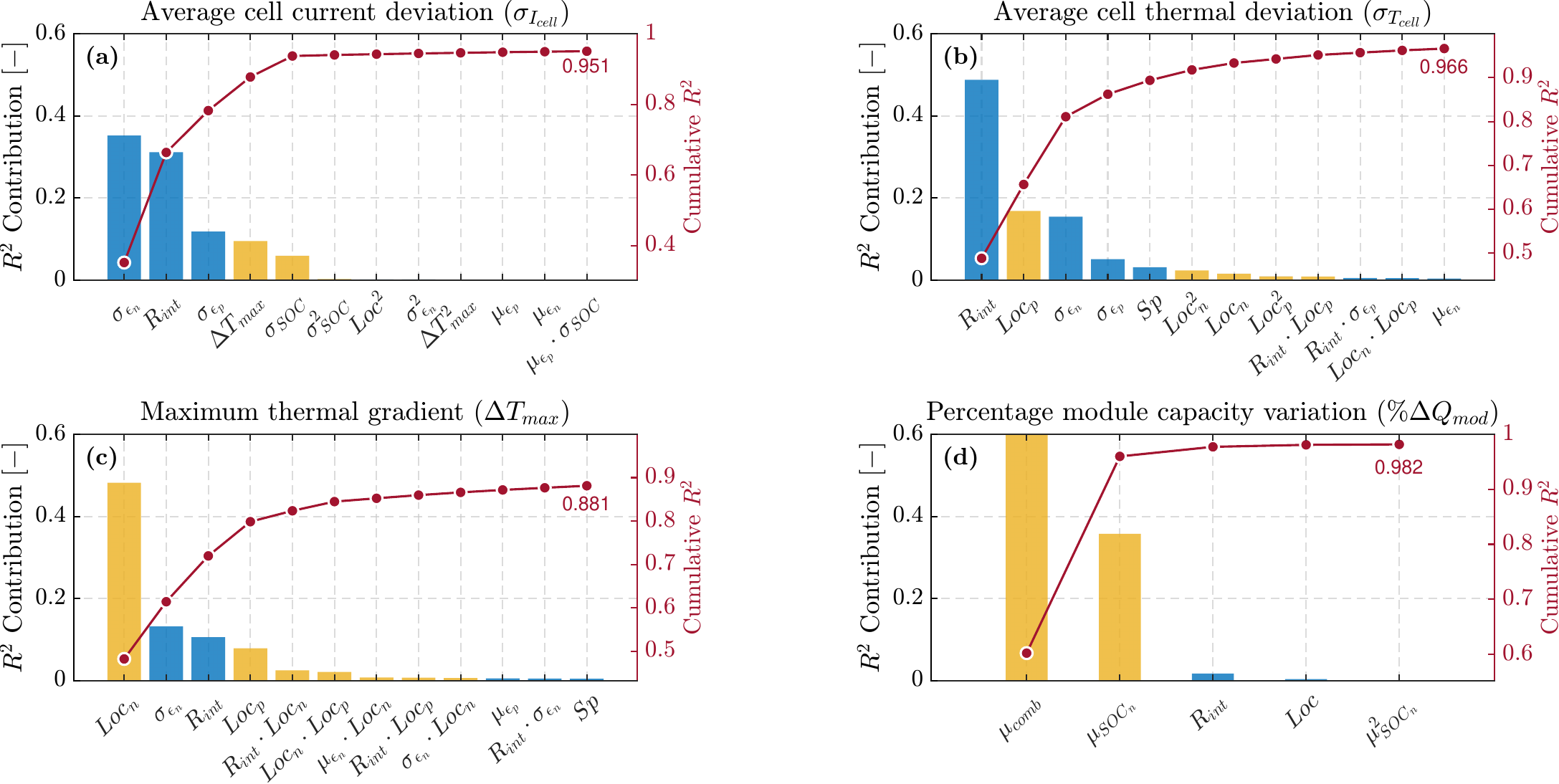}
  \includegraphics[trim=0 0 0 0,clip,width = 0.5\textwidth]{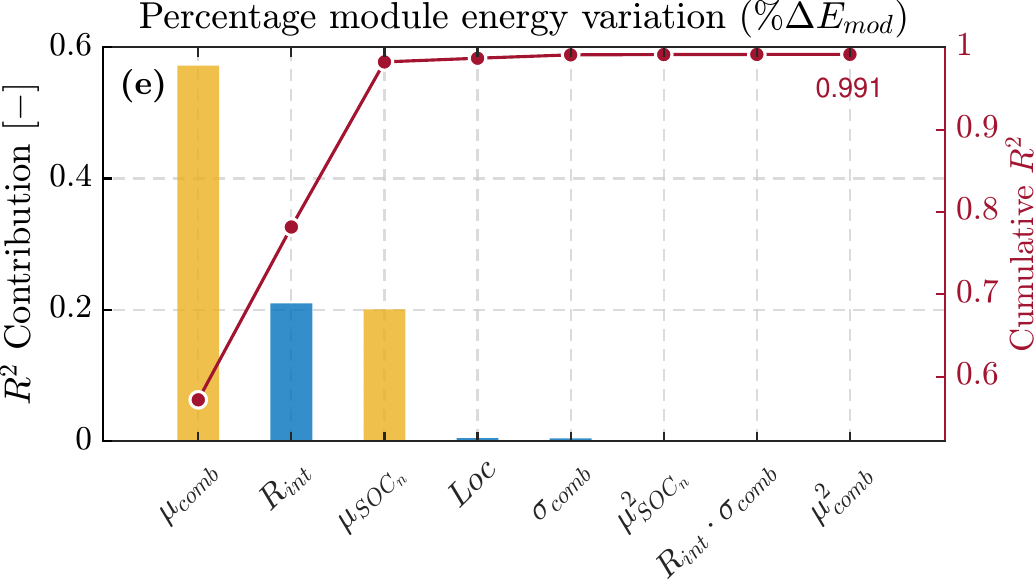}
  \caption{Figures (a-e) present the Pareto plot for each considered response variable in the short-term scenario. Compared to Figure \ref{fig:Pareto_1}, the MLR models are enhanced with ad hoc additional predictors (highlighted in dark yellow) to improve model fitting, as detailed in Section \ref{Sec:shortTerm_analysis}.}
  \label{fig:Pareto_2}
\end{figure}

\begin{enumerate}
\item \textbf{Current distribution standard deviation ($\sigma I_{cell}$):}
  $R_{int}$ and $\sigma_{\epsilon_{s,j}}$ (with $j \in [n,p]$) emerge as the primary drivers influencing the heterogeneous current distribution among the parallel branches, as shown in Figure \ref{fig:Pareto_1}(a). Specifically, the current gradient increases at the beginning of the operation as the $R_{int}$ value rises. In contrast, a higher $\sigma_{\epsilon_{s,j}}$ results in larger current heterogeneities in the middle and final SOC range.
  When cells with unmatched $\epsilon_{s,j}$ are connected, the electrode particles experience different lithium-ion intercalation and deintercalation fluxes during operation.
  This discrepancy leads to varying surface concentrations across the cells, causing them to operate at different $U_j$ and $\eta_j$. With the same overall current injected into the module, this imbalance in $U_j$ and $\eta_j$ results in an uneven distribution of current among the different branches, exacerbating current heterogeneities.
  Although the $R^2$ value of the model using the $\Gamma_1$ predictor set is satisfactory (i.e. 0.82), the model fitting can be further enhanced by including the $\Delta T_{max}$ and $\sigma SOC$ as additional predictors.
  A high thermal gradient causes significant variations in the temperature-dependent parameters within the cell-level ESPMs, particularly increasing heterogeneity in the particle diffusion coefficient among the cells. This exacerbates the differences in surface concentration ($c_{s,j}^{surf}$), further impacting the current distribution. 
  Similarly, large $\sigma_{SOC}$ indicates that the cells have operated at different SOC, resulting in dissimilar OCV values.

\item \textbf{Percentage variation in the  module capacity and energy ($\%\Delta Q_{mod}$ and $\%\Delta E_{mod}$):} The average value of the electrode active material volume fraction (\(\mu_{\epsilon_{s,j}}\) with \(j \in \{n, p\}\)), is the primary factor influencing the module overall capacity and energy, as shown in Figures \ref{fig:Pareto_1}(d) and (e), respectively. In a parallel connection scenario, the module capacity is the sum of the capacities of individual cells. Because the capacity of a single cell is directly proportional to \(\epsilon_{s,j}\) \cite{lee2019estimation}, a higher \(\mu_{\epsilon_{s,j}}\) (indicating greater \(\epsilon_{s,j}\) values among the cells) suggests an higher module total capacity, as each cell contribution is larger \cite{gong2014study}.
Furthermore, \(R_{\text{int}}\) plays a significant role in \(\%\Delta E\). As \(R_{\text{int}}\) increases, Joule losses rise, thereby reducing the module’s overall energy efficiency.

According to Figure \ref{fig:Pareto_1}(d-e), the \(R^2\) values below 0.7 for both MLR models indicate that the current predictors set do not adequately capture the data variability. To enhance the models fitting performance and gain a deeper understanding of the factors influencing module capacity/energy, we introduced two key modifications to the predictor list:
\begin{enumerate}
  \item \textbf{Lumped Mean and Standard Deviation of \(\epsilon_{s,j}\) (\(\mu_{comb}\) and \(\sigma_{comb}\)):}  
  The original predictors \(\mu_{\epsilon_{s,j}}\) and \(\sigma_{\epsilon_{s,j}}\) were replaced with \(\mu_{comb}\) and \(\sigma_{comb}\) to account for the capacity limitation imposed by the weakest electrode in each cell, calculated as in equation \eqref{eq:mu_comb}.
  \item \textbf{Average SOC of the $N_p$ cells at the end of the discharge cycle. (\(\mu SOC\)):}  
  We incorporated the mean state of charge (\(\mu SOC\)) as an additional predictor. A higher \(\mu SOC\) indicates that the energy extracted from the module during cycling is not entirely depleted upon reaching the cutoff voltage, meaning the cells remain slightly charged.
\end{enumerate}
The inclusion of \(\mu_{comb}\) and \(\mu_{SOC}\) significantly improved the models, as reflected in the \(R^2\) values, which increased dramatically from 0.57 and 0.70 to 0.98 and 0.99 for \(\%\Delta Q_{mod}\) and \(\%\Delta E_{mod}\), respectively (see Figure \ref{fig:Pareto_2})(d-e).

\item \textbf{Thermal distribution standard deviation ($\sigma T_{cell}$) and maximum thermal gradient ($\Delta T_{max}$):}
  Internal resistance ($R_{\text{int}}$) and the standard deviation of the active material volume fraction ($\sigma_{\epsilon_{s,j}}$, where $j \in \{n,p\}$) are the primary factors influencing heterogeneous thermal distribution among parallel cells, as shown in Figure \ref{fig:Pareto_1}(b-c). Notably, $R_{\text{int}}$ and $\sigma_{\epsilon_{s,j}}$ are also the dominant predictors of the standard deviation of cell current ($\sigma_{I_{cell}}$), confirming that increased unevenness in current distribution, especially in the absence of a thermal management system, leads to greater temperature heterogeneity. Although a similar trend is observed for $\Delta T_{max}$, the $R^2$ value of 0.42 indicates that the predictor set does not fully capture the response variability, thus, a key modification to the predictor list is introduced:
  \begin{enumerate}
    \item The Loc index is divided into $Loc_n$ and $Loc_p$ to specifically account for the perturbation positions of the negative and positive electrodes, respectively. The rationale behind this change is to track the positions of cells with the highest and lowest capacities in the negative electrodes. As will become clear in Section \ref{Sec:Arrangement}, $\Delta T_{max}$ occurs at low SOC values primarily due to variations in the anode OCP, leading to significant current gradients.
  \end{enumerate}
  According to Figures \ref{fig:Pareto_2}(b-c), the $R^2$ values increased significantly from 0.82 and 0.41 to 0.9 and 0.88 for $\sigma T_{cell}$ and $\sigma_{I_{cell}}$, respectively.
\end{enumerate}

\begin{figure}
  \centering
  \includegraphics[trim=0 0 0 0,clip,width = 1\textwidth]{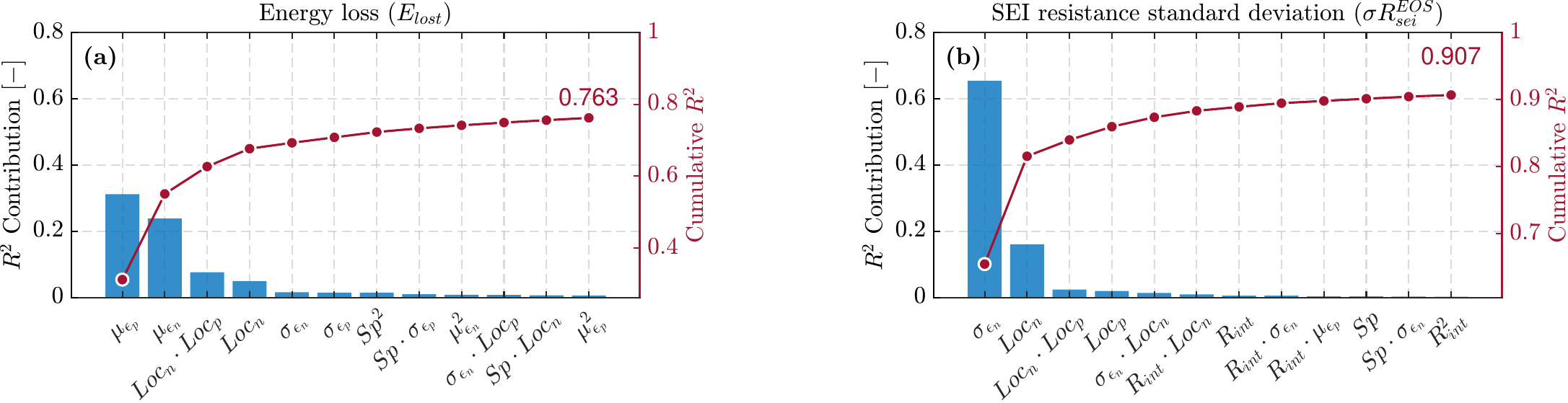}
  \caption{MLR analysis long-term simulation results. Figures (a) and (b) rank the terms of both MLR models based on their influence on the model $R^2$.}
  \label{fig:Fig_pareto_long}
\end{figure}

\subsubsection{MLR analysis: long term CtC heterogeneity effects} \label{Sec:longTerm_analysis}
The long-term impact of CtC variation on the performance and health of a battery module over 500 cycles is thoroughly investigated. This evaluation focuses on four critical responses: the energy loss between the beginning and end of the simulation ($E_{lost}$) and the standard deviation of the SEI resistance at the end of service life ($\sigma R_{SEI}^{EOS}$).
The resulting MLR models are reported in Figure \ref{fig:Fig_pareto_long},  which illustrates the contribution of each term in the models to the overall model $R^2$ for both $E_{lost}$ and $\sigma R_{SEI}^{EOS}$.
Overall, the results suggest that electrode-based manufacturing-induced heterogeneities significantly influence the aging behavior of parallel-connected battery modules. Specifically, the energy lost per cycle is higher in battery modules with a lower mean value of the active material volume fraction ($\mu_{\epsilon_{s,j}}$), as shown in Figure \ref{fig:Fig_pareto_long}(a). This occurs because, as discussed in the previous section, a higher mean $\mu_{\epsilon_{s,j}}$ leads to cells with greater capacity. These higher-capacity cells can handle more current, resulting in a lower real C-rate. Conversely, the standard deviation of the SEI resistance at the end of service life ($\sigma R_{SEI}^{EOS}$) strongly depends on the initial standard deviation of the active material volume fraction ($\sigma_{\epsilon_{s,n}}$), as illustrated in  Figure \ref{fig:Fig_pareto_long}(b). This indicates that initial heterogeneities play a crucial role in determining the variability of SEI resistance at the end of the battery's life.

The cell location emerges as the second most important factors.
As emphasized in Section \ref{Sec:shortTerm_analysis}, the primary reason for this outcome is the strong correlation between the locations of the negative and positive electrodes ($Loc_n$ and $Loc_p$) and the thermal gradient within the parallel-connected branches. These thermal gradients are responsible for triggering aging mechanisms within the system.
Indeed, our findings are consistent with previous experiments conducted in this field. For instance, \cite{liu2019effect} demonstrated that a 6 parallel-connected cells battery module subjected to a thermal gradient of $20^{\circ} C$ experienced a 5.2\% higher degradation rate compared to the isothermal case. This trend was further validated through experimental evidence by \cite{naylor2023battery}, who observed an accelerated aging rate in 2 parallel-connected cells exposed to a thermal gradient of $25^{\circ} C$. Furthermore, the authors highlighted that a high thermal gradient within the module led to divergent capacity fade among the parallel-connected cells.

\subsection{Cell arrangement strategy for thermal gradient reduction} \label{Sec:Arrangement}
According to the previous section, the degradation and propagation of heterogeneity in parallel-connected battery modules are significantly influenced by the initial heterogeneity characteristics of the fresh cells (i.e $\mu_{\epsilon_{s,j}}$ and $\sigma_{\epsilon_{s,j}}$). 
The $\mu_{\epsilon_{s,j}}$ and $\sigma_{\epsilon_{s,j}}$ of the initial cell parameters are primarily determined by the manufacturing process and cannot be optimized post-production. However, the arrangement of cells within the module, which emerged as the second most critical predictors, provides an opportunity to enhance the overall performance and longevity of the battery module.

This study introduces a straightforward cell arrangement strategy for a parallel-connected battery module.
The proposed strategy involves positioning cells with larger capacities at the beginning of the module, where the interconnection resistance is lower due to the proximity to the module terminals. 
A lower $R_{int}$ results in higher delivered currents during the initial phase of operation. By locating the cells with the highest capacity at the beginning of the module, it can effectively manage higher currents and mitigate SOC differences during operation.
This arrangement results in a lower thermal gradient at low SOC levels, as the cells experience more uniform SOC.
Practically, the proposed cell arrangement strategy involves characterizing the individual cells before building the module. Then, the cells with higher capacity should be placed in locations where the interconnection resistance is expected to be lower. In situations where budget and/or time constraints make the preliminary single-cell testing infeasible, a sub-optimal alternative could be arranging the parallel cells according to their weight. In \cite{an2016rate}, a linear correlation between cell capacity and weight was demonstrated for a batch of 5300 5.3Ah cells. 

\begin{figure}[H]
  \centering
  \includegraphics[trim=0 0 0 0,clip,width=1\linewidth]{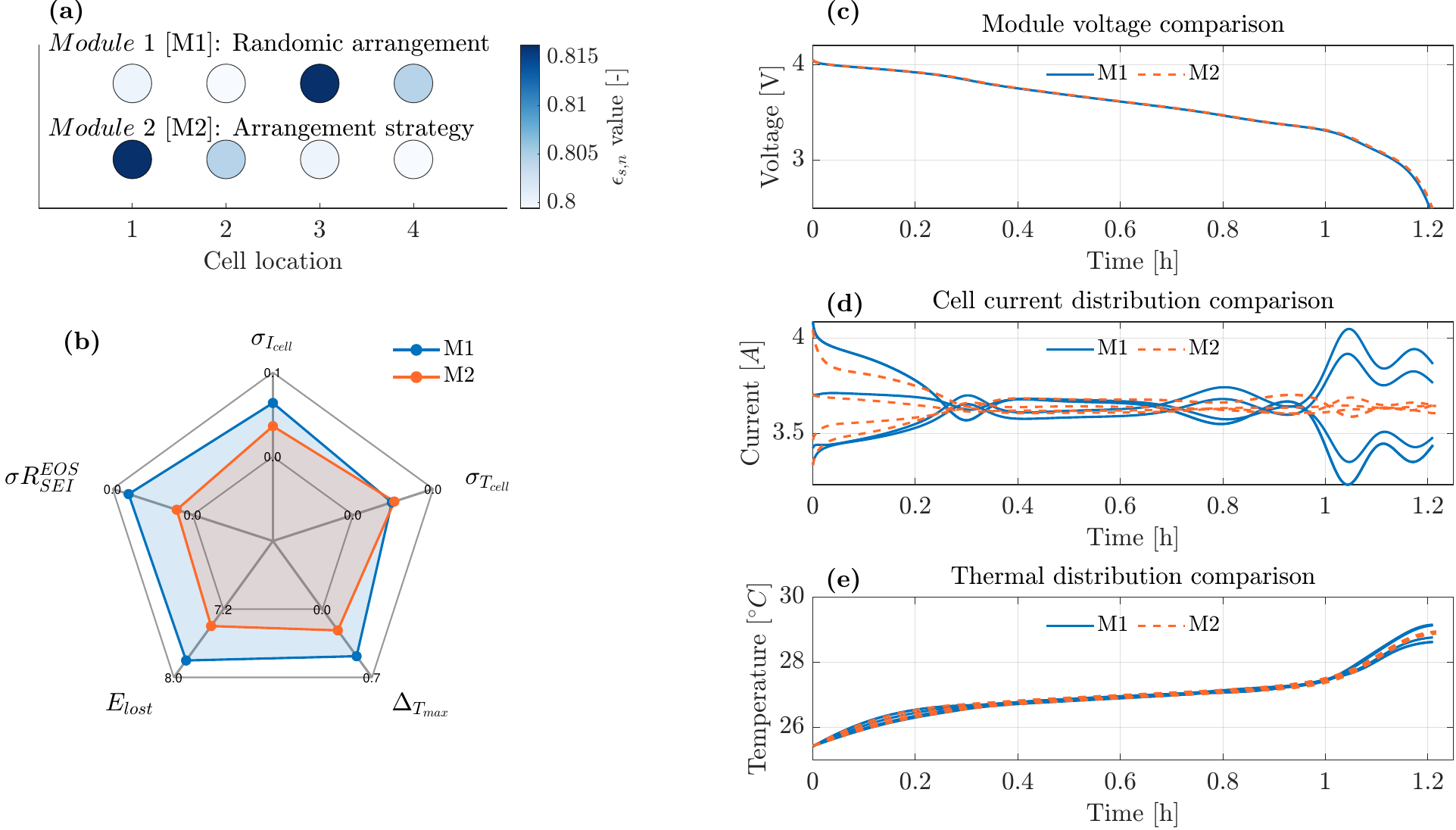}
  \caption{Cell arrangement strategy for parallel-connected module graphical overview. (a) shows the battery module with randomly arranged cells (M1) and the optimized cell arranged(M2).  (b) compare the modules' response variables. (c), (d), and (e) report the module voltage, the cell current distribution, and the thermal distribution comparison between M1 and M2, respectively.}
  \label{fig:Cell_arrang}
\end{figure}

A visual representation of the cell arrangement strategy is provided in Figure \ref{fig:Cell_arrang}. To demonstrate this approach, a 4-cell battery module (M1) is randomly selected from the MLR analysis simulations conducted in the previous section. 
Figure \ref{fig:Cell_arrang}(a) also visualizes the rearranged module (M2). It is noticeable that cells with a higher active material volume fraction (i.e., higher capacity) are arranged in ascending order, from the highest, positioned near the module terminal, to the lowest in the last position.
The thermal and current distribution of M1 is compared with that of M2, which is achieved by rearranging the cells in descending order based on their capacity, in Figure \ref{fig:Cell_arrang}(d) and (e), respectively.
The cell with the highest capacity delivered the largest current at high SOC and effectively managed higher currents, mitigating large SOC gradients, resulting in homogeneous current and thermal distribution toward the end of the cycle. Specifically, $\sigma I_{cell}$ and $\Delta T_{max}$ decrease from 0.0649 A and 0.471$^{\circ} C$ to 0.0573 A and 0.227$^{\circ} C$, as highlighted in Figure \ref{fig:Cell_arrang}(b). This results in a reduction of 5.2\% (from -7.83\% to -7.42\%) and 60.9\% (from 2.4643e-06 to 9.641e-07) in $E_{lost}$ and $\sigma R_{SEI}^{EOS}$, respectively.
Overall, the benefits of the arrangement strategy, such as decreased energy loss and improved uniformity in SEI resistance, demonstrate its effectiveness, especially in achieving more uniform aged battery modules for second-life applications \cite{piombo2024novel, moy2024second}.

\section{Conclusion} \label{Sec:Conclusion}
This study assessed the impact of heterogeneity among parallel cells, resulting from manufacturing tolerances and various module configurations, on the module's performance.
To the best of our knowledge, this article is the first to consider this type of analysis for parallel cells. Additionally, unlike other studies relying on empirical evaluations, our approach employs a model-based method to carry out the analysis and interpret the resulting performance.
Leveraging the high interpretability and low computational demands of multi-linear regression models, the analysis was conducted focusing on the identified key cell-level parameters, as well as various model configurations. The objective was to assess their relative importance concerning overall capacity, energy, heterogeneity distribution, and aging propagation in a parallel-connected module.
In the short-term MLR analysis, electrode active material volume fractions and the interconnection resistance significantly impact module performance and the propagation of heterogeneities.  Modules with high overall capacity and energy tend to have a higher average value of single-cell capacities, as highlighted by the effect of the predictors $\mu_{comb}$ on the responses $\%\Delta E$ and $\%\Delta Q$. Conversely, the standard deviation in the capacities of interconnected cells strongly affects both current distribution and thermal inconsistencies within the module.
In the long-term MLR analysis, it is demonstrated that a high thermal gradient accelerates the module's aging rate and increases the SEI resistance variation across interconnected cells.
It is worth noting that favoring more uniformly aged cells at the end of their first life is desirable also for creating durable and secure second-life battery modules and packs.
To address this, a simple cell arrangement strategy is presented to reduce the CtC thermal gradient, thereby decreasing aging heterogeneities in the long-term scenario. The key idea is to leverage the cell current resulting from the manufacturing-induced CtC variation to mitigate the module's thermal gradient by placing cells with higher capacity at the beginning of the module, where the interconnection resistance is lower.
From an implementation perspective, a cost-effective solution could involve arranging the cells based on their weight.
According to simulation results, this method allows for a thermal gradient reduction of 51.8\%  and a consequent decrease of 5.2\%  of the module energy loss after 500 aging cycles.
In summary, our findings demonstrate the importance of considering electrode manufacturing-dependent parameters and the potential benefits of the proposed cell arrangement strategy in enhancing the performance, safety, and longevity of parallel-connected battery modules.

\section*{Limitations and further work}
This study presents several limitations that should be considered. 
Specifically, the experiments and high-fidelity simulations were conducted at a fixed mild C-rate (i.e., below 1C), which is representative of the maximum discharging rate operating conditions \cite{pozzato2023analysis}. However, it is recognized that variations in C-rate, especially at higher charge/discharge rates, can significantly affect cell behavior. To improve the generalizability of the findings across different operational conditions, future research should incorporate higher C-rate scenarios.
Additionally, different ambient temperature conditions could be considered to further assess the temperature-dependent behavior of the cells and modules. It is important to note that when 
extreme operating conditions, such as aggressive current profiles or high ambient temperatures, can lead to aging mechanisms like lithium plating and particle cracking \cite{edge2021lithium}. These phenomena should be considered in the model, as demonstrated in previous studies \cite{ai2022composite, o2022lithium, pozzato2021modeling}.
Future work should broaden the study by incorporating different cell architectures (e.g., cylindrical, prismatic, pouch), chemistries (e.g., LFP, LMO), and form factors (e.g., 18650, 4680), as well as more realistic load profiles based on real-world cycling conditions.

\section*{Author contributions}
\textbf{Simone Fasolato:} Conceptualization, Methodology, Software, Validation, Formal analysis, Investigation, Data curation, Visualization, Writing - original draft, Writing - Review \& Editing. 
\textbf{Anirudh Allam:} Methodology, Software.
\textbf{Simona Onori:} Supervision, Methodology, Funding acquisition, Writing - review \& editing. 
\textbf{Davide M. Raimondo:} Supervision, Conceptualization, Methodology, Visualization, Funding acquisition, Writing - original draft, Writing - review \& editing.

\section*{Competing Interests}
The authors declare that there is no non-competing financial interests or personal relationships that could have appeared to influence the work reporter in this paper. 

\section*{Data availability}
\begin{itemize}
  \item Both cell- and module-level experimental data used in this work are available at \cite{piombo2024full}.
  \item The simulation dataset used for the Multi-linear regression analysis are generated implementing the battery module model described in the Method section in MatLab R2021b. 
  All the simulations were performed on a Windows 11 personal computer with 16 Gbytes of RAM and a 2.8 GHz 11th Gen Intel(R) Core i7 processor. The Model ODEs were solved using CasADi \cite{andersson2012casadi}, a symbolic framework for automatic differentiation.
\end{itemize}

\section*{Computer code}
The model code and post-processing codes that have been used to produce the results of this study are available upon reasonable request.

\section*{Acknowledgement}
The authors gratefully acknowledge the support of the Stanford Energy Control Lab for hosting the experimental activities. The research presented within this paper was partially supported by the Bits and Watts Initiative and StorageX Initiative within the Precourt Institute for Energy at Stanford University and by the CHIPS Joint Undertaking and the European Union through the EcoMobility project (grant 101112306).

\bibliographystyle{IEEEtran}
\bibliography{biblio} 

\end{document}